\documentclass[aps,prf,showpacs,floats,twocolumn,floats,superscriptaddress,floatfix,english]{revtex4-1}
\usepackage{graphicx}
\usepackage{array}
\usepackage{multirow}
\usepackage{setspace}

\usepackage{multirow}
\usepackage{caption}
\usepackage{color}
\usepackage{epstopdf}
\usepackage{epsfig}
\usepackage[normalem]{ulem}
\usepackage{babel}
\usepackage{float}
\usepackage{amsmath}

\def\tb{\textcolor{black}}

\def\be{\begin{equation}}
\def\ee{\end{equation}}

\def\Hovmoller{H\"ovm\"oller }

\def\RBC{Rayleigh-B\'enard convection }
\def\RBCalt{Rayleigh-B\'enard convection}
\def\RRBC{rotating \RBC}
\def\RRBCalt{rotating \RBCalt}
\def\Racb{Ra_c^\mathrm{\tiny{bulk}}}
\def\RacwD{Ra_{cw}^\mathrm{\tiny{FT}}}
\def\RacwN{Ra_{cw}^\mathrm{\tiny{ZF}}}
\def\Racone{Ra_c^{(1)}}
\def\Ractwo{Ra_c^{(2)}}
\def\thcone{\theta_c^{(1)}}
\def\thctwo{\theta_c^{(2)}}

\begin{document}

\title{Prograde and meandering wall modes in rotating Rayleigh-B\'enard convection with conducting walls}

\author{S. Ravichandran}
\affiliation{Nordita, KTH Royal Institute of Technology and Stockholm University, Stockholm 10691, Sweden}
\affiliation{Centre for Climate Studies, Indian Institute of Technology Bombay, Mumbai 400076, India}
\author{J. S. Wettlaufer}
\affiliation{Nordita, KTH Royal Institute of Technology and Stockholm University, Stockholm 10691, Sweden}
\affiliation{Yale University, New Haven, Connecticut 06520-8109, USA}

\makeatother
\begin{abstract}
{We use direct numerical simulations to study convection} in \RRBC in horizontally confined geometries of a given aspect ratio, with the walls held at fixed temperatures. We show that this arrangement is unconditionally unstable to flow that takes the form of wall-adjacent convection rolls. For wall temperatures close to the temperatures of the upper or lower boundaries, we show that the base state undergoes a Hopf bifurcation to a state comprised of spatio-temporal oscillations--`wall modes'--precessing in a retrograde direction. {We study the saturated nonlinear state of these modes, and} show that the velocity boundary conditions at the upper and lower boundaries are crucial to the formation and propagation of the wall modes: asymmetric velocity boundary conditions at the upper and lower boundaries can lead to prograde wall modes, while stress-free boundary conditions at both walls can lead to wall modes that have no preferred direction of propagation.
\end{abstract}
\maketitle

\section{Introduction}

Rotating convection is an essential process in, among many other systems, stars \citep{Sreeni2020}, protoplanetary disks \citep{Gerbig2022}, planetary interiors \citep{Aurnou2015,Bercovici2015, Aurnou2020}, Earth's atmosphere and oceans \citep[e.g.][]{Emanuel1994,Marshall1999,Griffiths2022}, and the cryosphere \citep[e.g.][]{Wells2008,Ravichandran2021a,Claudia2023}. Moreover, rotating \RBC acts as a model system for these phenomena, and the study of pattern formation, such as transient axisymmetric rings observed during spin-up \citep[e.g.,][]{Boubnov1986,Vorobieff1998,Zhong2010,Ravichandran2020_RotCon}, and travelling waves observed near the onset of convection \citep{Ecke1992,Ning1993,liuEckhausBenjaminFeirInstabilityRotating1997a,liuNonlinearTravelingWaves1999a}. In typical model studies of \RRBCalt, both theory and numerical simulations employ horizontally periodic domains, whereas experiments do not, leading in part to disagreements. For instance, \cite{Rossby1969} observed that in \RRBC  with the typical no-slip experimental boundary conditions, the onset of convection occurs for significantly smaller thermal forcing than predicted by the linear stability analysis of \cite{Chandrasekhar1953}.  Although explanations of this apparent disagreement between theory and experiment can be traced to the basic differences between boundary conditions and finite amplitude perturbations \cite[e.g.,][]{Veronis1966b, Veronis1968, Herrmann1993, Kuo1993}, there are many interesting outstanding questions \citep[See][]{Ecke2023}. For example, measurements suggest that the peripheral modes may also be responsible for the mismatch between the Nusselt numbers in laboratory experiments and numerical simulations \citep{Kunnen2013,Ecke2022,DeWit2020}.

In the absence of rotation, the onset of convection in a horizontally unbounded layer of fluid of depth $H$ across which a temperature difference $\Delta T$ is maintained, is governed by the dimensionless Rayleigh number
\be
Ra = \frac{g\alpha\Delta T H^3}{\nu \kappa}, \label{eq:Ra}
\ee
where $g$ is the acceleration of gravity and $\nu$ and $\kappa$ are the viscosity and thermal diffusivity of the fluid. The onset of convection occurs when a critical Rayleigh number, $\Racb$, is exceeded, where $\Racb$ = $O(10^3)$, with the exact value depending on the boundary conditions. \cite{Chandrasekhar1953} showed that, independent of the Prandtl number $Pr=\nu/\kappa$, the instability that leads to convection is non-oscillatory. 

Rotation about the vertical axis suppresses the effects of buoyancy and thus enhances stability.  Therefore, the critical Rayleigh number increases with the rotation rate, $\Omega$, as
\be
\Racb \simeq  E^{-4/3} , \label{eq:Rac_bulk}
\ee
where $E=\nu / 2 \Omega H^2$ is the Ekman number \citep[][]{Chandrasekhar1953,Veronis1966b}. In contrast to non-rotating systems, the onset of convection in horizontally unbounded \RRBC can be oscillatory if the Prandtl number $Pr<0.69$ \citep{Chandrasekhar1953}. 

\citet{Rossby1969} showed experimentally that convection sets in for much smaller Rayleigh numbers than predicted by Eq. (\ref{eq:Rac_bulk}), the possible origins of which were discussed contemporaneously by \citet[][]{Veronis1968} as being associated with boundary conditions. Of relevance to our study, the experiments by \cite{Ecke1992}, supported by linear stability analysis by \cite{Goldstein1993}, showed that the wall-adjacent convection takes the form of a travelling wave with a phase speed opposite to the sense of rotation. They also showed that the system undergoes a Hopf bifurcation at a critical Rayleigh number, following which the travelling wave appears. 

\cite{Herrmann1993} and \cite{Kuo1993} independently showed that, in the asymptotic limit of $E\rightarrow 0$, the critical Rayleigh number for the onset of the travelling waves in confined \RRBC with adiabatic walls is 
\be
\RacwN = \pi^2 (6 \sqrt{3} )^{1/2} E^{-1} = 31.82 E^{-1}, \label{eq:RacwN}
\ee
where the superscript ZF denotes `zero flux'.
\cite{Herrmann1993} also showed that conducting walls stabilize the wall modes, with a critical Rayleigh number that has the same leading order scaling as that for an infinite layer, but with a smaller prefactor:
\be
\RacwD = 0.9086 \left(\pi E^{-1} \right)^{4/3} + 2.124 \left(\pi E^{-1} \right)^{7/6} = 4.18 E^{-4/3} + 8.08 E^{-7/6}, \label{eq:RacwD}
\ee
where the superscript FT denotes `fixed temperature' highlighting the fact that mathematical well-posedness requires the temperature to be prescribed at a conducting boundary. In their analysis, \cite{Herrmann1993} assume a wall temperature equal to the purely conductive (linear) profile
\be
\bar{T}_{\mathrm{\tiny{wall}}} = 1-z. \label{eq:lin_temp}
\ee
The flow, for $Ra > \RacwD$, again takes the form of travelling waves that propagate in a retrograde sense, against the sense of rotation.

Instead of the linear temperature boundary condition of \cite{Herrmann1993}, if a constant temperature is imposed across the entire height of the walls, the system is unconditionally unstable with the flow taking the form of wall-adjacent convective rolls. {Similar rolls were observed by \cite{ning1993spatial} for weak thermal forcing at the walls.} \tb{Constant wall temperatures may be relevant in the melting of ocean-terminating glaciers due to the rotation-influenced convection, wherein the coupling between convective structures and melting morphology could become important \citep[][]{Ravichandran2021a}.}

{For supercritical Rayleigh numbers, the wall modes attain a nonlinear saturated steady state consisting of wall-adjacent regions of upwelling and downwelling flow propagating in a retrograde direction. This nonlinear state has been observed by \cite{lopezOnsetConvectionModerate2007, Favier2020,Zhang2020}, and \cite{DeWit2020}, and \cite{Favier2020} showed it to be robust to severe non-axisymmetric modifications of the geometry. For Rayleigh numbers well beyond the onset of bulk convection, the nonlinear state becomes the so-called boundary zonal flow \citep[BZF, see e.g.][]{Zhang2020,DeWit2020,Ecke2022,wedi_experimental_2022}, which has been shown to be responsible for significant amounts of heat transfer in \RRBC \citep[]{DeWit2020}.}

{Motivated by these findings of the influence of the boundary conditions on the dynamical state of wall modes, here we examine the effects of changing the uniform temperature at which the walls and upper and lower boundaries are held, along with the velocity boundary conditions (BC), as summarized in Table \ref{tab:BCs}. We show that the steady roll state undergoes a Hopf bifurcation as a function of the wall temperature, leading to wall modes. The wall temperature at which this onset occurs must be found from linear stability analysis, with the steady rolls as the base state.  We study the nonlinear state of the resulting instability, comparing it with the nonlinear state of wall modes with adiabatic walls. We find that the velocity BCs at the upper and lower boundaries control the direction of propagation of the wall modes, and prograde wall modes can arise for suitable velocity BCs. Finally, by studying both cuboidal and cylindrical geometries, we confirm the findings of \cite{Favier2020} that the nonlinear wall mode state is robust to non-axisymmetric geometric modifications.} 

The rest of the paper is organized as follows. {In \S \ref{sec:setup}, we describe the geometry of the problem and the numerical method used for the simulations.} In \S \ref{sec:results}, we present results from the numerical simulations as the wall temperature, {the velocity BC at the upper boundary, the horizontal cross-section of the domain, and the Prandtl number} are varied, compare our results with other known travelling wave solutions in \RRBCalt, and discuss the effects of asymmetric velocity BCs at the upper and lower boundaries on the direction of propagation of the wall-adjacent spatio-temporal patterns. We conclude in \S \ref{sec:conclusion}.

\section{Setup and Numerical Simulations}\label{sec:setup}
\begin{figure}
\centering
\includegraphics[width=0.5\columnwidth]{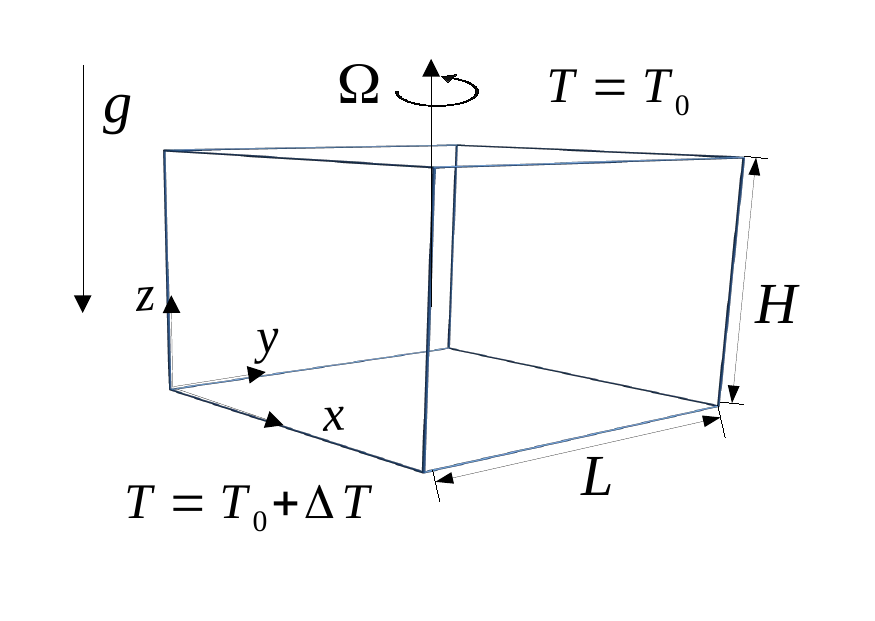}
\caption{{Our simulations are performed in a cuboidal volume of square cross-section and aspect ratio $L/H=2$, {so that the domain is the region $|x|\leq1$, $|y|\leq1$ $0\leq z \leq 1$.} The walls are either adiabatic or held at a constant temperature. The container rotates about the vertical axis with a constant angular velocity, with gravity pointing vertically down.}\label{fig:schematic}}
\end{figure}

The domain is a rectangular volume of height $H$ and width $L$, with an aspect ratio $\Gamma=L/H=2$, shown schematically in figure \ref{fig:schematic}. The system rotates about the vertical $z-$axis with a constant angular velocity $\Omega$. 

We make the Boussinesq approximation, so that fluid properties are assumed to be constant and the flow is assumed to be incompressible.
We nondimensionalize the governing equations using the length scale $H$ and the buoyancy velocity scale $U_b=(g\alpha \Delta T H)^{1/2}$. The nondimensional governing equations become
\begin{align}
\frac{D\boldsymbol{u}}{Dt} & =-\nabla p - \frac{\sqrt{Pr}}{E\cdot \sqrt{Ra}} \boldsymbol{e}_z \times \boldsymbol{u}+ \left(\frac{Pr}{Ra}\right)^{1/2}\nabla^{2}\boldsymbol{u}+\boldsymbol{e}_{z}\theta,\label{eq:momentum}\\
\nabla\cdot\boldsymbol{u} & =0, \label{eq:continuity}\\
\frac{D\theta}{Dt} & = \left(\frac{1}{Ra Pr}\right)^{1/2}\nabla^{2}\theta.\label{eq:temperature}
\end{align}

We consider the combinations of BCs listed in Table \ref{tab:BCs}.
\begin{widetext}
\begin{table}
    \centering
    \begin{tabular}{c|c|c|c}
    BC & $z=0$ & $z=1$ & Walls \\
    \hline
    SYMNS &  $u = v = w = 0$, $\theta=1$ & $u = v = w = 0$, $\theta=0$ &  $u = v = w = 0$, $\theta=\theta_w$ or $\partial \theta / \partial n = 0$   \\
    ASYM &  $u = v = w = 0$, $\theta=1$ & $w=0, \partial (u,v)/\partial z = 0$, $\theta=0$ &  $u = v = w = 0$,  $\theta=\theta_w$ or $\partial \theta / \partial n = 0$  \\
    \end{tabular}
    \caption{\label{tab:BCs} The combinations of boundary conditions (BCs) considered. Symmetric (`SYMNS') BCs have no-slip upper and lower boundaries, while asymmetric (`ASYM') BCs have \tb{stress-free} upper and no-slip lower boundaries. Consequences of symmetric \tb{stress-free} (`SYMFS') BCs, defined in analogy with SYMNS BCs, and of \tb{stress-free} BCs at all boundaries (`ALLFS') are examined in \S \ref{sec:other_BCs}.}
\end{table}
\end{widetext}
The temperature of the lower boundary is fixed at $\theta(z=0)=1$, the upper boundary at $\theta(z=1)=0$, and the walls are either adiabatic ($\partial \theta / \partial n = 0$) or have a fixed temperature $\theta=\theta_w$, with $0\leq\theta_w\leq1$. The velocity obeys either no-slip or stress-free BCs on the walls at $x=\pm1,y=\pm1$, and the lower and upper boundaries at $z=0$ and $z=1$ respectively.

Subject to these BCs, equations (\ref{eq:momentum}--\ref{eq:temperature}) are solved using the finite volume solver \emph{Megha}-5, used in previous studies of convection \citep[][]{Ravichandran2020_RotCon,Ravichandran2021a,Ravichandran2021b}. The solver uses second-order central differences in space and a second-order Adams-Bashforth timestepping scheme. {Simulations are initialized with broadband noise added to the initial conditions which trigger convection.} {We use a grid resolution of up to $256^3$ uniformly spaced points in the three space directions and a timestep $dt \geq 1.25\times10^{-3}$. The Nusselt number changes by only a few percent, with no change in the convection pattern, when the vertical resolution is changed from $128$ to $256$ grid points; and by approximately $0.1\%$ when the horizontal resolution is changed from $256$ to $512$ grid points. }

\section{Results and Discussion}\label{sec:results}

{The governing dimensionless parameters of Equations (\ref{eq:momentum}--\ref{eq:temperature}) are the Ekman ($E$), Rayleigh ($Ra$), and Prandtl ($Pr$), numbers as defined above. Simulations are run for a given aspect ratio $\Gamma$, and set of boundary conditions. We minimize the amount of computation required by exploring the effects of varying the parameters one at a time around the point $E=10^{-4}$, $Ra=10^6$, $Pr=1$, with $\Gamma=2$, and $0\leq\theta_w\leq1$. For this combination of $E$ and $Pr$ in a horizontally unbounded geometry, $\Racb\approx1.5\times10^6$, whereas with insulating walls wall modes appear for $Ra > \RacwN=3.2\times10^5$. For walls with the linear temperature profile (Eq. \ref{eq:lin_temp}), $\RacwD=1.3\times10^6$, and for $Ra > \RacwD$ retrograde propagating wall modes are obtained regardless of the asymmetry in the velocity BCs. This is discussed further in \S \ref{sec:velocities}.}

{The initial and boundary conditions described in \S \ref{sec:setup} lead to the onset of convection everywhere in the domain.  The associated bulk flow structure decays away, leaving only the wall attached convection, from which the wall modes emerge and grow into their nonlinear state.  The process from which the wall modes emerge from the wall attached convective state involves an instability of a transient base state, and the simulations show a robust and rapid growth into the nonlinear state.  Thus, we focus on the latter situation and leave the stability analysis of the secular base state for a stand alone study.}

\subsection{Symmetric and asymmetric velocity boundary conditions} \label{sec:SYMNS_ASYM}

\begin{figure}
\centering
\includegraphics[width=0.8\columnwidth,trim={50 0 25 0},clip]{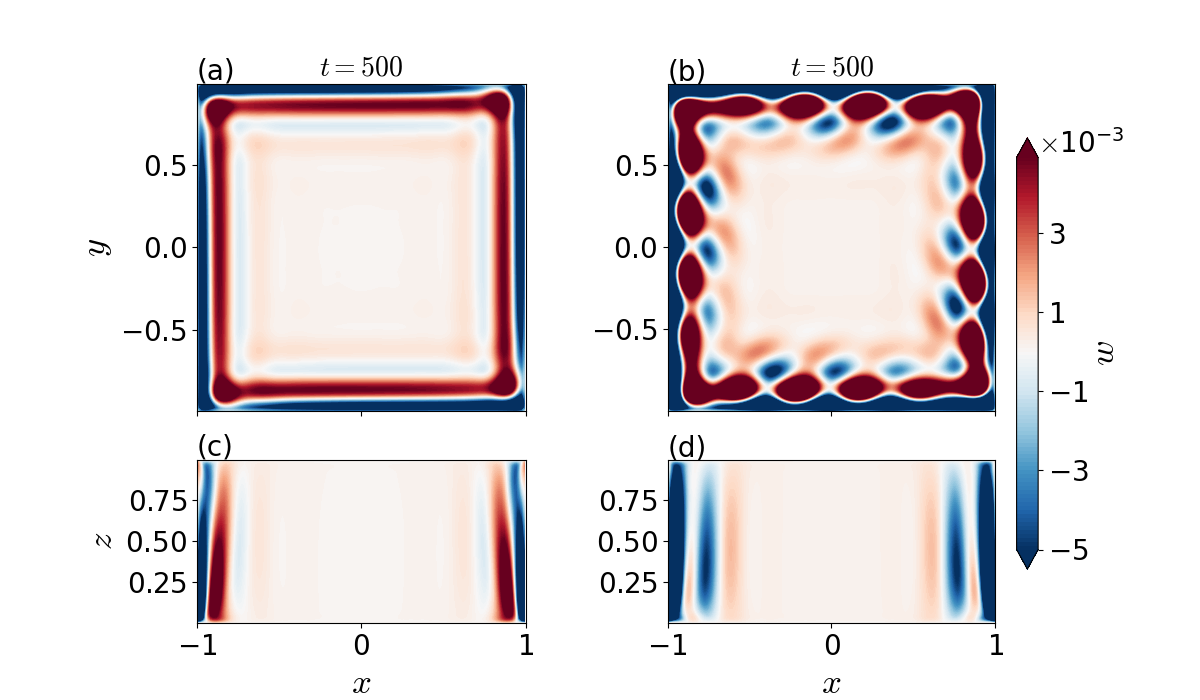}
\caption{\label{fig:wall_modes_symns} With $E=10^{-4}$, $Ra=10^6$, $Pr=1$ and SYMNS BCs, we see a steady convective state with wall-adjacent rolls for (a,c) $\theta_w=0.25$; and wall modes for (b,d) $\theta_w=0.1$. The subplots show the horizontal (a,b) and vertical (c,d) cross-sections of the vertical velocity $w$.}
\end{figure}

\begin{figure}
\centering
\includegraphics[width=\columnwidth]{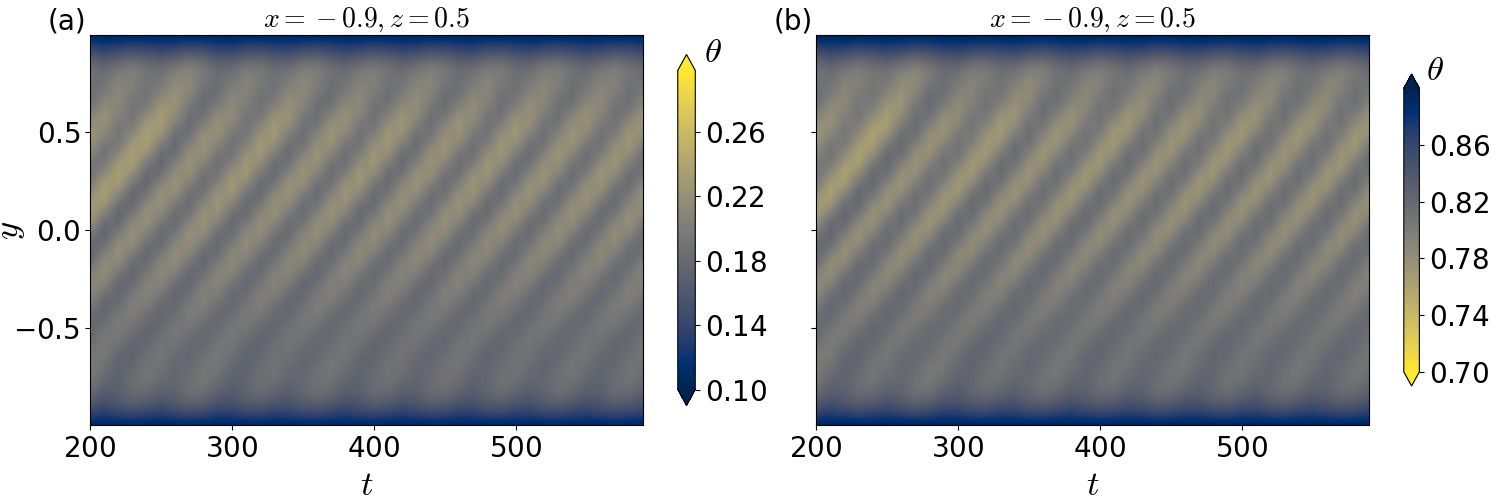}
\caption{\label{fig:Hovmoller_SYMNS} \tb{Space-time} \Hovmoller diagrams of the temperature $\theta$ for $E=10^{-4}, Ra=10^6, Pr=1$, and (a) $\theta_w=0.1$, and (b) $\theta_w=0.9$. The slope of the patterns shows that the wall modes propagate in a retrograde direction. The symmetry of the Boussinesq equations ensures that the flow patterns for $\theta_w$ and $1-\theta_w$ are similar. {Note that the color scheme is inverted in (b).}}
\end{figure}

We first examine the flow structures that arise with SYMNS BCs (see Table \ref{tab:BCs}) for $E=10^{-4}, Ra=10^6, Pr=1, \Gamma=2$, while varying $\theta_w$. Figure \ref{fig:wall_modes_symns} shows that two types of flow may arise; a wall temperature of $\theta_w = 0.25$ leads to steady wall-attached rolls in the shape of the container, and $\theta_w=0.1$ generates retrograde wall modes. {The former are similar to the time-averaged flow seen in the experiments of \cite{ning1993spatial} at supercritical Rayleigh numbers, with weak thermal forcing at the imperfectly conducting walls.} 
The retrograde propagation of the wall modes is apparent from the \tb{space-time} \Hovmoller diagram in figure \ref{fig:Hovmoller_SYMNS}(a), where we show the near-wall temperature along a horizontal line. Due to the symmetry of the problem, retrograde wall modes with a similar structure are observed for $1-\theta_w=0.1$, as seen in figure \ref{fig:Hovmoller_SYMNS}(b). \tb{Note that the downward vertical velocity seen in the wall-adjacent region for $\theta_w < 0.5$ (Fig. \ref{fig:wall_modes_symns}) would be upward were $\theta_w > 0.5$.}

Wall modes are also seen with ASYM BCs for wall temperatures close to upper or lower boundary values, as shown in figures \ref{fig:wall_modes_asym}. 
{In figure \ref{fig:Hovmoller_ASYM}, we plot the \Hovmoller diagrams for the space-time evolution of the wall modes, showing a reversal of the direction of propagation as follows.  For small $\theta_w = 0.05$, we find retrograde wall modes, as shown in figure \ref{fig:Hovmoller_ASYM}(a), whereas for large $\theta_w = 1$, the wall modes travel in the prograde direction, as shown in figure \ref{fig:Hovmoller_ASYM}(b).  
\begin{figure}
\centering
\includegraphics[width=0.8\columnwidth,trim={50 0 25 0},clip]{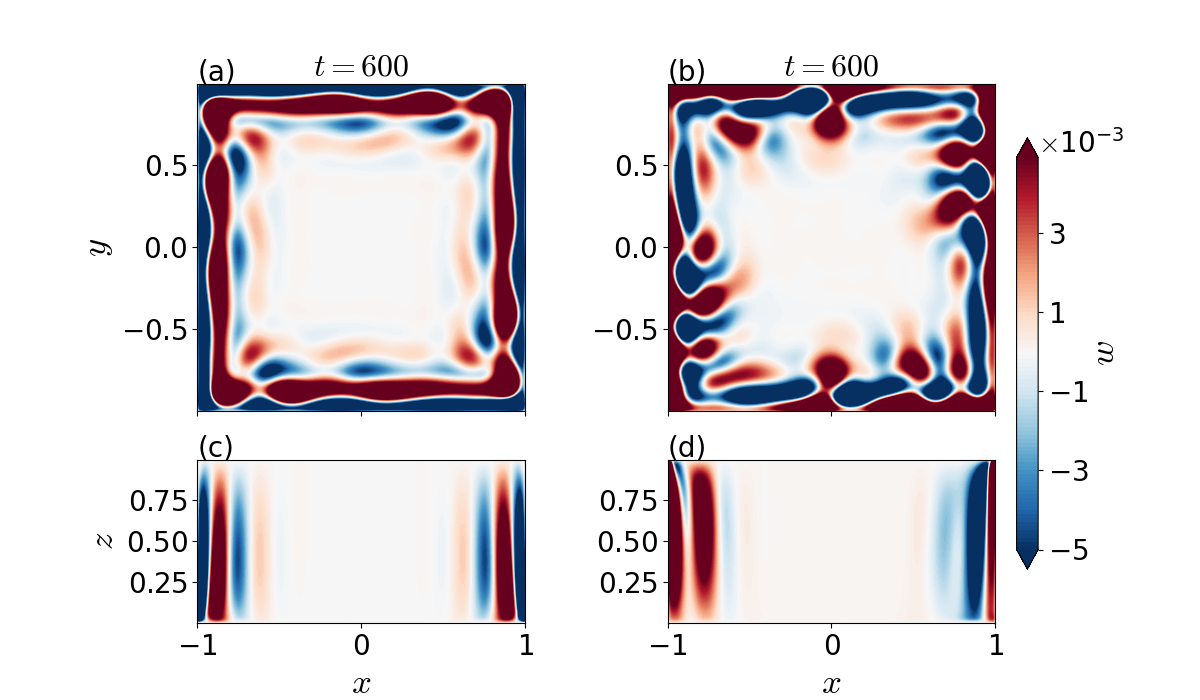}
\caption{\label{fig:wall_modes_asym} The wall modes observed for $E=10^{-4}$, $Pr=1$, $Ra=10^6$ and ASYM BCs for (a,c) $\theta_w=0.05$ and (b,d) $\theta_w=1$. The subplots show the horizontal \tb{(a,b)} and vertical \tb{(c,d)} cross-sections of the vertical velocity $w$. In (a,c), the wall modes propagate in the retrograde direction, while in (b,d) they propagate in the prograde direction (see figure \ref{fig:Hovmoller_ASYM}). }
\end{figure}
Finally, for intermediate $\theta_w$ we find convection in steady rolls, as was the case with SYMNS BCs.}
\begin{figure}
\centering
\includegraphics[width=\columnwidth]{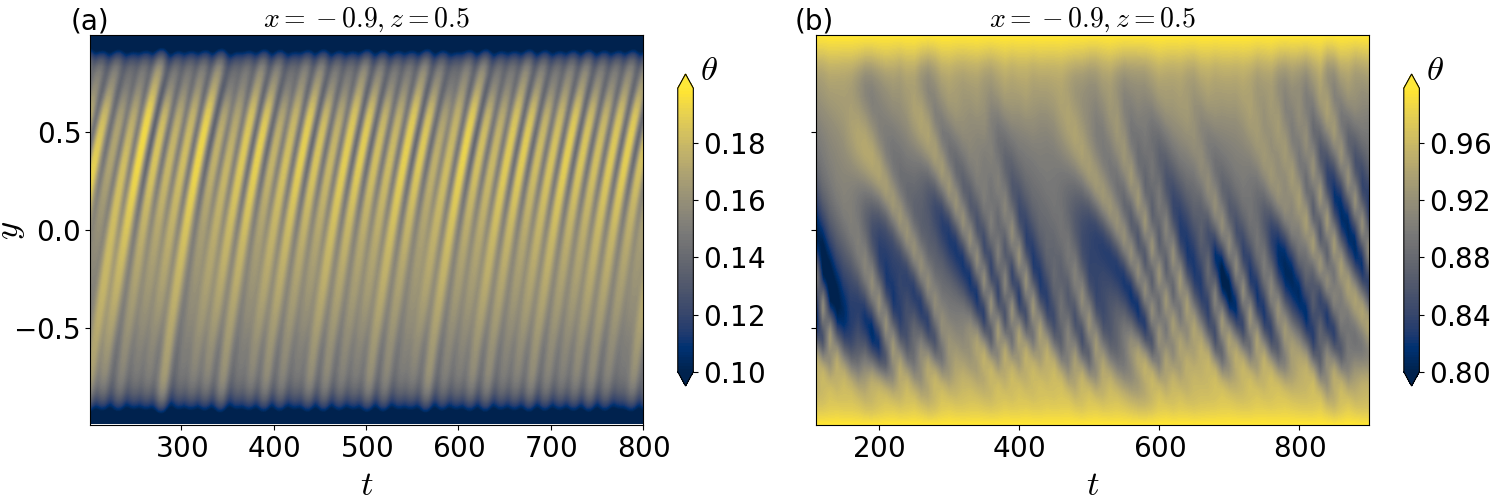}
\caption{\label{fig:Hovmoller_ASYM} \Hovmoller diagrams of the temperature $\theta$ for the same parameters as in figures \ref{fig:wall_modes_asym}, showing that for (a) $\theta_w=0.05$ the wall modes travel in the retrograde direction, while for (b) $\theta_w=1$ they travel in the prograde direction.}
\end{figure}

\cite{Horn2017} find flow features that propagate in the prograde direction for small aspect ratios ($\Gamma=0.5$) \tb{with precession freqencies that are comparable to retrograde modes}. However, these flow features are `bulk modes', and not restricted to the near-wall region. \tb{\cite{Horn2017} associate their bulk modes } with the slow modes of \cite{Goldstein1993}, with precession frequencies  much smaller than those of the retrograde modes. Neither of these studies considers the effect of asymmetric velocity BCs, although \cite{Goldstein1993} do consider conducting walls.
Here, we see prograde \emph{wall modes} with asymmetric velocity BCs, but only for $\theta_w \rightarrow 1$, \tb{with precession frequencies that are comparable to the usual retrograde modes}. In contrast, for insulating walls, the wall modes are retrograde even for asymmetric BCs; and if the wall temperature is very different from that of the no-slip boundary, say $\theta_w \lesssim 0.9$, the flow is comprised of nested rolls. 
{In \S \ref{sec:velocities}, we explain this behavior by examining the tangential velocity at the walls.}

\subsection{Additional combinations of boundary conditions} \label{sec:other_BCs}

\begin{figure}
\centering
\includegraphics[width=0.8\columnwidth,trim={50 0 25 0},clip]{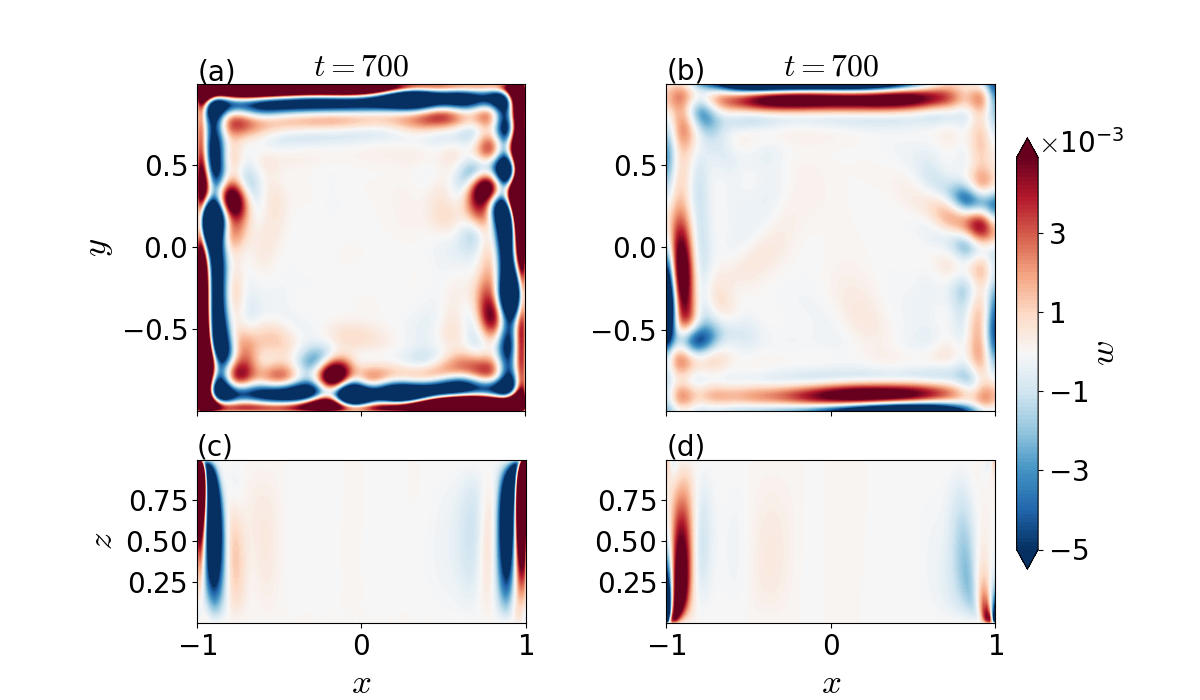}
\caption{\label{fig:wall_modes_misc_bcs} The wall modes observed for $E=10^{-4}$, $Ra=10^6$, $Pr=1$ and (a,c) SYMFS BCs with $\theta_w=0.95$ and (b,d) ALLFS BCs with $\theta_w=0.1$. The subplots show the horizontal (a,b) and vertical (c,d) cross-sections of the vertical velocity $w$. The wall modes have no fixed direction of propagation (see also figure \ref{fig:Hovmoller_misc_bcs}) {and the strong four-fold symmetry seen in Figs. \ref{fig:wall_modes_symns} and \ref{fig:wall_modes_asym} is lost}.}
\end{figure}

\begin{figure}
\centering
\includegraphics[width=\columnwidth]{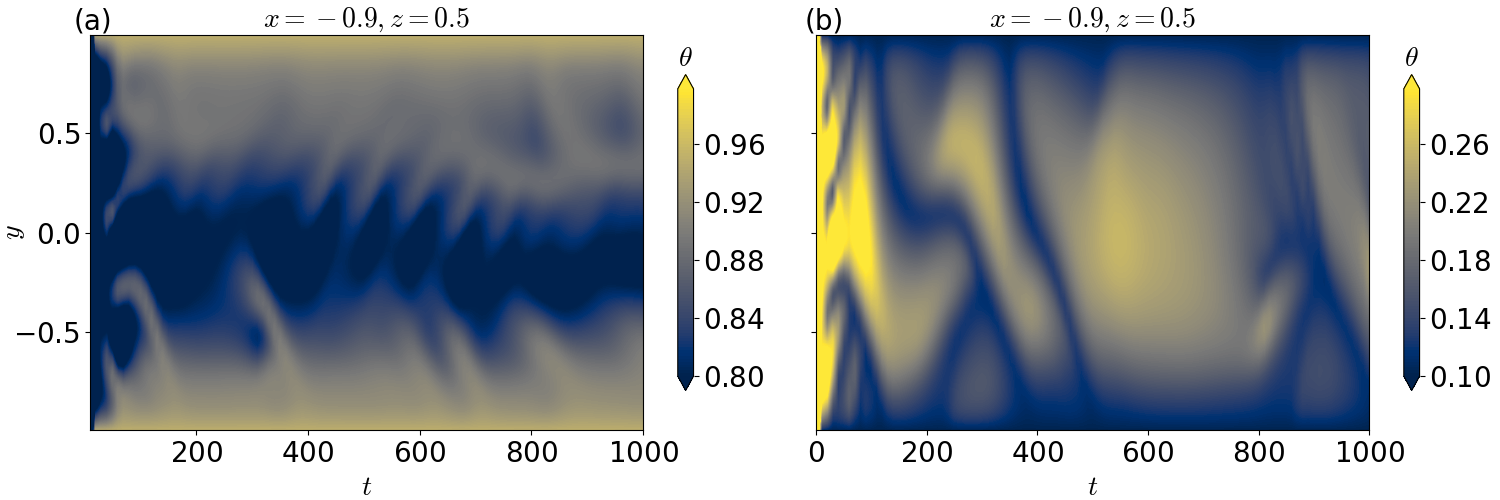}
\caption{\label{fig:Hovmoller_misc_bcs} \Hovmoller diagrams of the temperature $\theta$ for the same parameters as in figures \ref{fig:wall_modes_misc_bcs}, showing that for both (a) $\theta_w=0.95$ and SYMFS BCs, and (b) $\theta_w = 0.1$ and ALLFS BCs, the wall modes show no fixed direction of propagation.}
\end{figure}

If both the upper and lower boundaries are stress-free, and the walls are conducting and obey the no-slip condition (`SYMFS' BCs, see Table \ref{tab:BCs}), wall modes form but they have no clearly discernible direction of propagation. Insulating walls lead to the standard retrograde wall modes, in agreement with earlier studies. Similarly, if both the upper and lower boundaries as well as the conducting walls are stress-free (`ALLFS' BCs), the wall modes that appear have no fixed direction of propagation. Representative snapshots of the wall modes that result from SYMFS and ALLFS BCs are shown in figures \ref{fig:wall_modes_misc_bcs}(a) and (b) respectively, with the corresponding \Hovmoller diagrams shown in figures \ref{fig:Hovmoller_misc_bcs}(a) and (b) respectively. 
A comparison of these figures to the equivalent figures for SYMNS (figures \ref{fig:wall_modes_symns} and \ref{fig:Hovmoller_SYMNS}) and ASYM BCs (figures \ref{fig:wall_modes_asym} and \ref{fig:Hovmoller_ASYM}) suggests that while wall modes are observed with conducting walls for suitable wall temperatures $\theta_w$, no-slip velocity BCs on at least the upper or lower boundaries are necessary for wall modes to propagate in a definite direction. {We explain this behavior  by examining the tangential velocity at the walls in \S \ref{sec:velocities}.}  {Finally, we note that the flow in figure \ref{fig:wall_modes_misc_bcs} has lost the strong four-fold symmetric structure shown in figures \ref{fig:wall_modes_symns} and \ref{fig:wall_modes_asym} for SYMNS and ASYM BCs respectively.}

\subsection{{Cylindrical geometry}} \label{sec:cylinder}
\tb{The robustness of wall modes to changes in geometry, observed in the experiments of \cite{Ning1993} (but not published; see \cite{Ecke2023}), was studied numerically by \cite{Favier2020}.} \cite{terrien_suppression_2023} observed that wall modes can be suppressed by obstacles (`fins') on the boundary \tb{by increasing the thermal forcing required for wall mode onset}. Here we show the consistency of the flow structures between cuboidal and cylindrical geometries. \tb{We also find that the sharp corners in the cuboidal geometry act to dampen the travelling wall modes.}

{We show the geometry-independence of our findings by performing simulations in a cylindrical geometry of aspect ratio $D/H=2$ where $D$ is the diameter of the cylinder. The cylindrical geometry is embedded in the Cartesian geometry using volume penalization \citep[see][]{Ravichandran2020_RotCon}, with volume penalization parameters of $\eta=5\times10^{-3}$ or $\eta=10^{-3}$ giving the same results.  The SYM, ASYM and SYMFS BCs are defined in analogy with those in \S \ref{sec:setup} for the cuboidal geometry, whereas the ALLFS BC cannot be implemented with the solver used here. All other parameters, $E=10^{-4}$, $Pr=1$, $Ra=10^6$, are unchanged.}

\begin{figure}
\centering
\includegraphics[width=0.8\columnwidth,trim={50 0 10 0},clip]{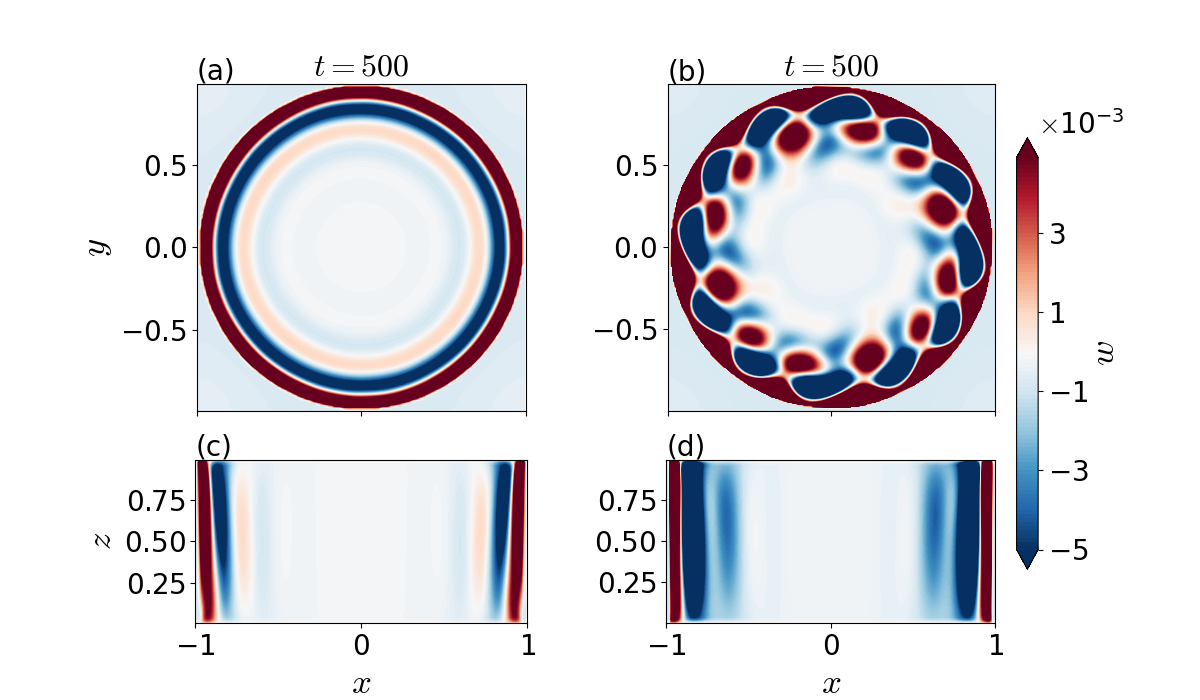}
\caption{\label{fig:cyl_wall_modes_symns} {With $E=10^{-4}$, $Ra=10^6$, $Pr=1$ and SYMNS BCs in the cylindrical geometry, we see a steady convective state with wall-adjacent rolls for (a,c) $\theta_w=0.8$; and wall modes for (b,d) $\theta_w=0.95$. The subplots show the horizontal (a,b) and vertical (c,d) cross-sections of the vertical velocity $w$.}}
\end{figure}

\begin{figure}
\centering
\includegraphics[width=\columnwidth]{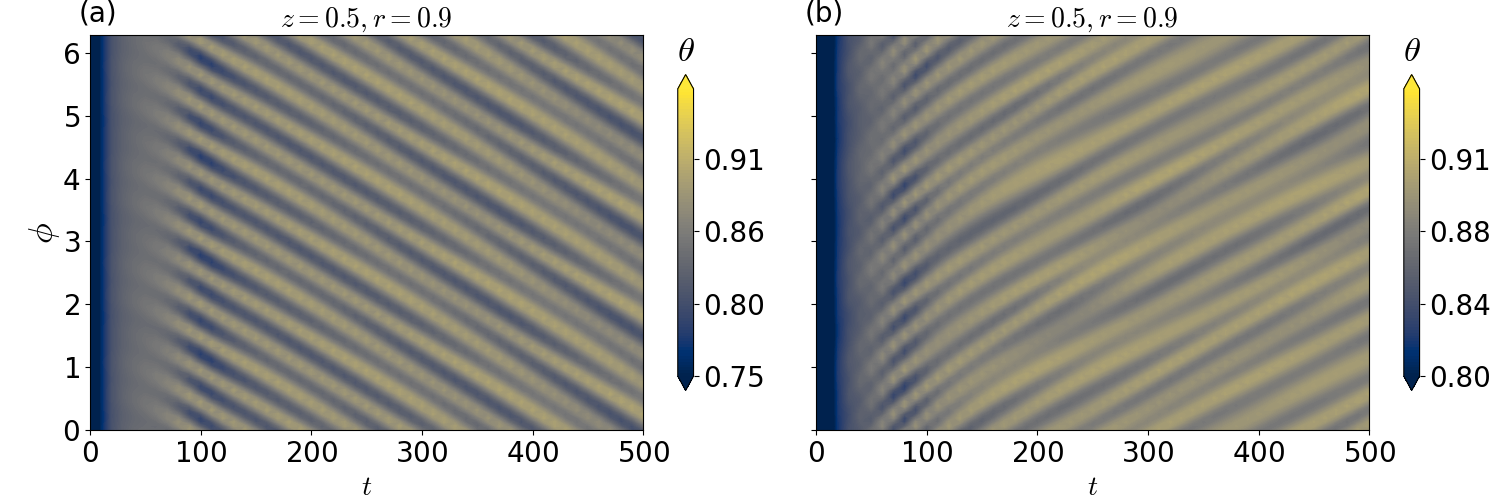}
\caption{\label{fig:Hovmoller_cyl} {\Hovmoller plots of the temperature $\theta$ for $E=10^{-4}$, $Pr=1$, $Ra=10^6$ and $\theta_w = 0.95$ in the cylindrical geometry shown in Fig. \ref{fig:cyl_wall_modes_symns}. The direction of propagation is (a) retrograde for SYM BCs and (b) prograde for ASYM BCs.}}
\end{figure}

{In Fig. \ref{fig:cyl_wall_modes_symns}, we show the steady roll state and retrograde precessing wall modes, which should be compared to those in Fig. \ref{fig:wall_modes_symns}. The principal quantitative difference is that the wall temperature $\thcone$ up to which wall modes are sustained is larger in the cylindrical geometry for the same Rayleigh number. These wall modes travel in a retrograde or prograde direction for SYM and ASYM BCs respectively. This is seen in the \Hovmoller diagrams in Fig. \ref{fig:Hovmoller_cyl}, for SYMNS and ASYMFS BCs respectively, showing the influence of asymmetric velocity BCs. }

\subsection{{Tangential flow velocity at the wall}} \label{sec:velocities}

{The reversal in the travel direction of the wall modes may be explained by examining the flow velocity tangential to the walls. The following arguments apply to both the cylindrical and cuboidal geometries.}

\begin{figure}
\centering
\includegraphics[width=\columnwidth]{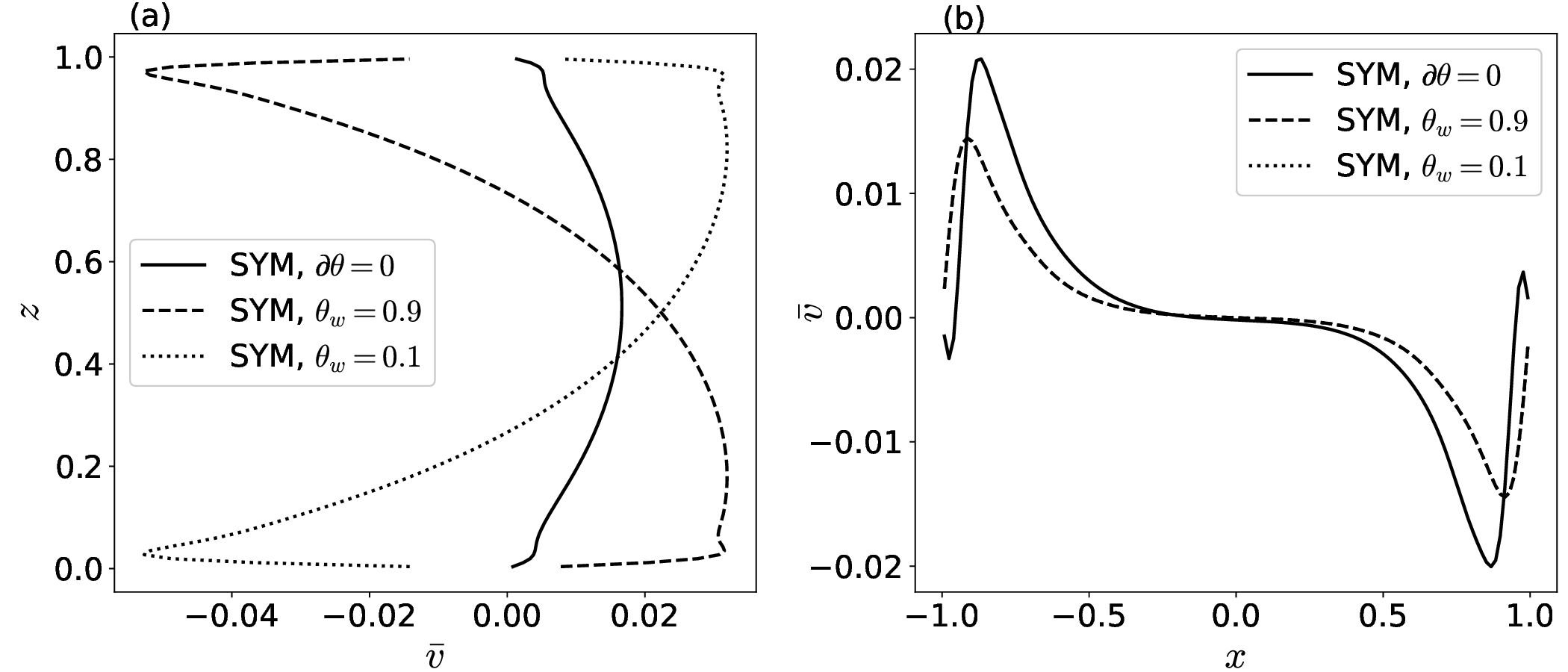}
\caption{\label{fig:tangvel_SYMNS} {With  $E=10^{-4}$, $Ra=10^6$, $Pr=1$ and SYM velocity BCs, (a) the time-averaged velocity $\bar{v}$ tangential to the wall at $x=-1$. The velocities are averaged over the region $x<-0.9, -0.75<y<1.25$ and over at least $300$ flow time units once the wall modes have set in. Large negative (thus prograde) velocities are seen near the upper and lower boundaries for $\theta_w=0.9$ and $\theta_w=0.1$ respectively. (b) The tangential velocity $\bar{v}(x)$ averaged over the region $-0.75 < y < 1.25, 0<z<1$. Despite the stark asymmetry in the vertical profiles of velocity for $\theta_w=0.1$ and $\theta_w=0.9$, the depth averaged velocity profiles are identical. The positive, and thus retrograde, average tangential velocities are consistent with the retrograde precession of the wall modes for these velocity BCs.}}
\end{figure}

{For SYM velocity BCs, the time-averaged tangential velocity $\bar{v}$ is plotted in Fig. \ref{fig:tangvel_SYMNS} for insulating and fixed wall temperatures. Insulating wall BCs give a vertically symmetric tangential velocity that vanishes at the upper and lower boundaries \citep[see also][]{Zhang2020}. The velocity profiles for conducting walls are starkly different.}

{Consider the flow at the wall with $\theta_w=0.9$. Whereas for the insulating case, there is an inner, O($E^{-1/4}$), Stewartson boundary layer with a prograde velocity, when the walls are at fixed temperature this boundary layer is absent and the velocities at the wall are entirely retrograde. The near-wall flow has a positive vertical velocity and turns inwards at the upper boundary, acquiring a prograde tangential velocity. Flow towards the wall at the lower boundary acquires a retrograde tangential velocity. The stronger buoyancy-forcing at the upper boundary, due to the larger wall-normal thermal gradient, leads to a vertical shear, or thermal wind, resulting in the skewed velocity profiles seen in Fig. \ref{fig:tangvel_SYMNS}(a). These arguments apply when $\theta_w=0.1$. }

\begin{figure}
\centering
\includegraphics[width=0.48\columnwidth]{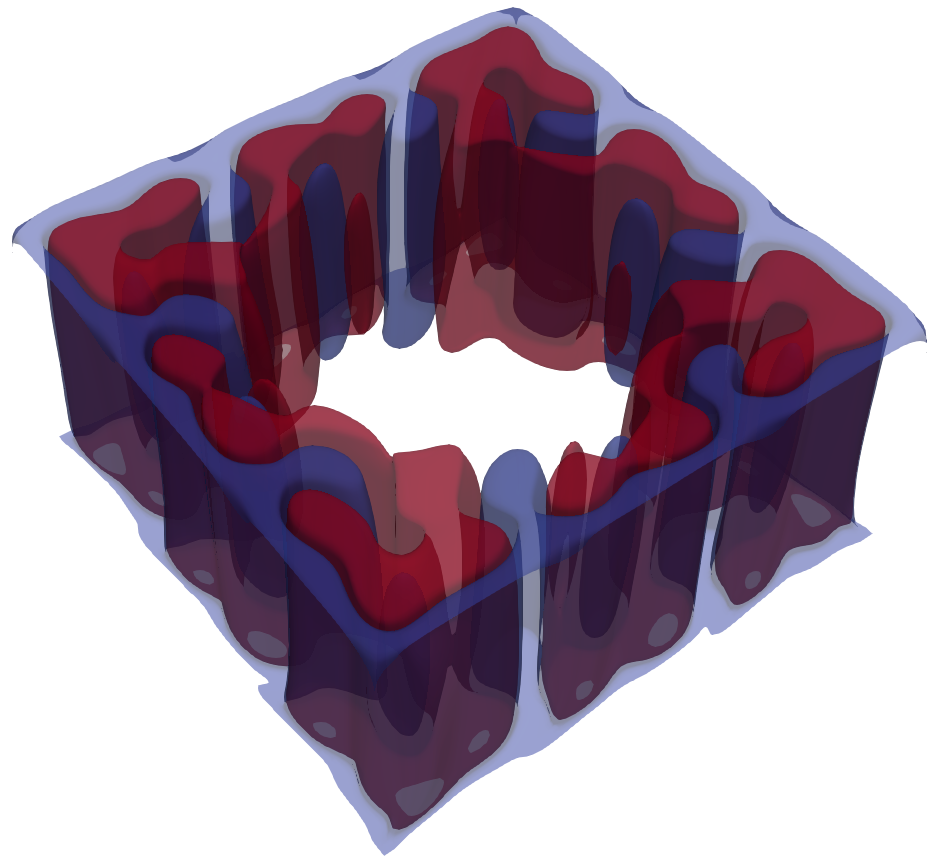}
\includegraphics[width=0.48\columnwidth]{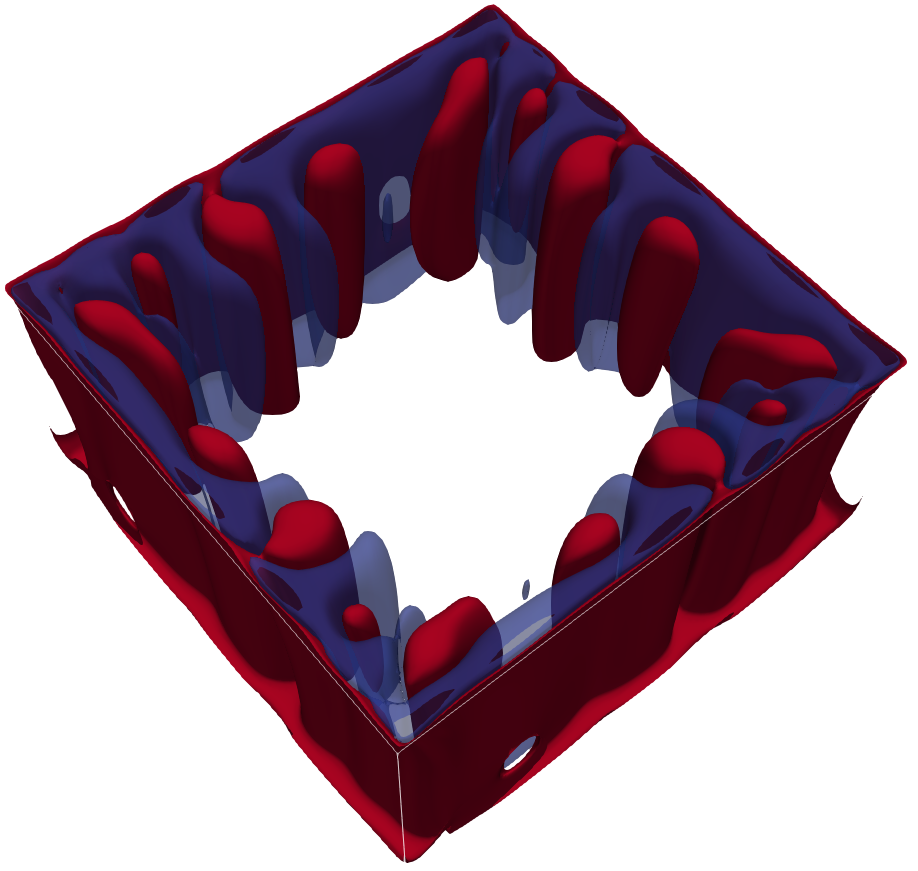}
\caption{\label{fig:3D_isocontours} {Three-dimensional isocontours of the vertical velocity $w$ for $E=10^{-4}$, $Pr=1$, $Ra=10^6$ and (a) $\theta_w=0.05$ with SYMNS BCs, and (b) $\theta_w=0.975$ with ASYM BCs. In both cases, it is evident that the flow structures are columnar, and hence rotation-dominated. The iso-contours are plotted for $w=0.001$ (red) and $w=-0.001$ (blue).}}
\end{figure}

{Despite the strong vertical shear seen in Fig. \ref{fig:tangvel_SYMNS}(a), the depth-averaged tangential velocity is retrograde, as shown in Fig. \ref{fig:tangvel_SYMNS}(b). The profiles for $\theta_w=0.1$ and $\theta_w=0.9$, expected to be similar by symmetry, have the same sense as in the case with insulating BCs. In the rotation-dominated flows considered here, wall modes remain vertically coherent, as clearly seen in the 3D contours of Fig. \ref{fig:3D_isocontours} for both SYM and ASYM BCs, and their precession direction is determined by the depth-averaged tangential velocity, as shown in Fig. \ref{fig:tangvel_SYMNS}(b). Thus, the precession direction of the wall modes is the same for $\theta_w=0.1$ and $0.9$, and insulating BCs.}

\begin{figure}
\centering
\includegraphics[width=1\columnwidth]{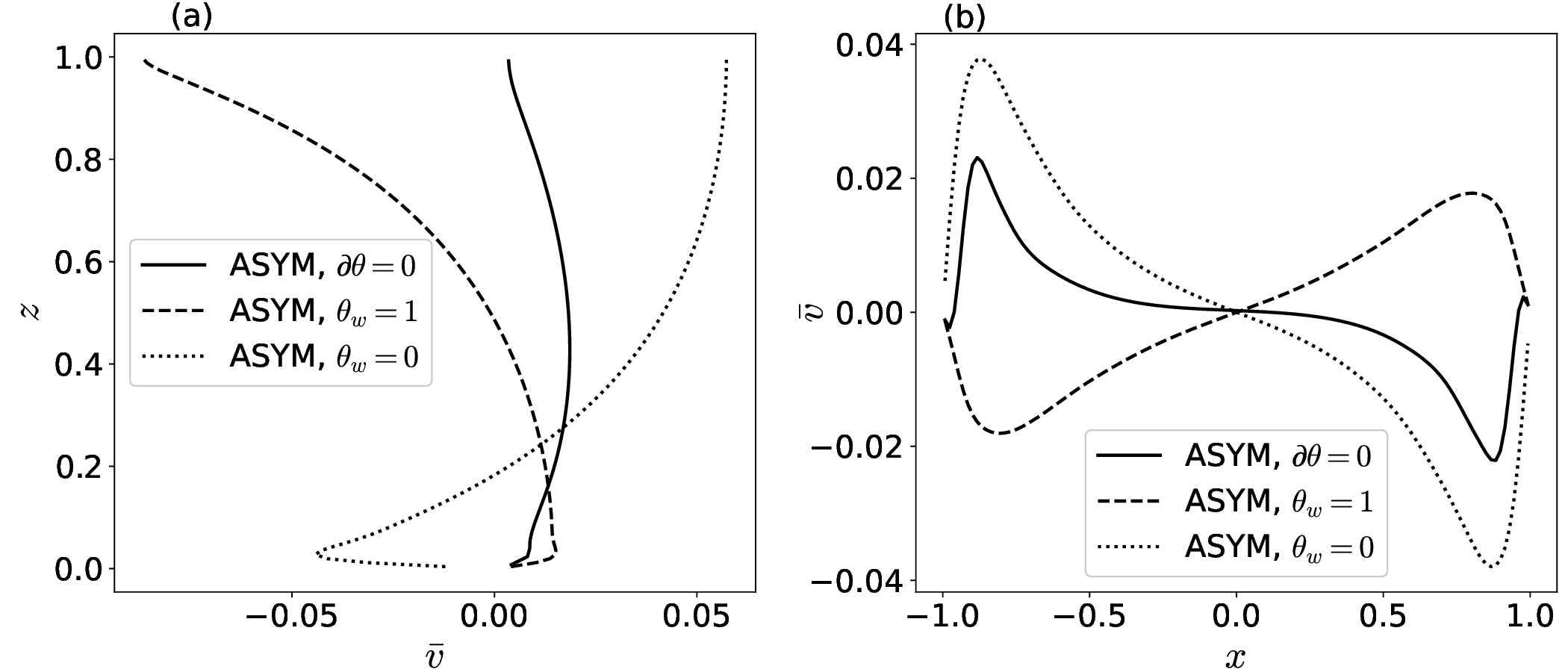}
\caption{\label{fig:tangvel_ASYM} {As in Figure \ref{fig:tangvel_SYMNS}, but with ASYM velocity BCs, showing the tangential velocity $\bar{v}$ as a function of (a) $z$ and (b) $x$.}}
\end{figure}

{The effects of ASYM BCs are shown in Fig. \ref{fig:tangvel_ASYM}(a,b), where the vertical profiles for $\theta_w=0$ and $\theta_w=1$ are no longer symmetric about $z=0.5$. Owing to the horizontal thermal gradient, the geostrophically balanced flow develops a vertical shear, where the sign of the shear depends on the direction of the thermal gradient. As a result, the depth-averaged velocity profile, which determines the direction of precession, is retrograde for $\theta_w=0$ and prograde for $\theta_w=1$. }

\begin{figure}
\centering
\includegraphics[width=1\columnwidth]{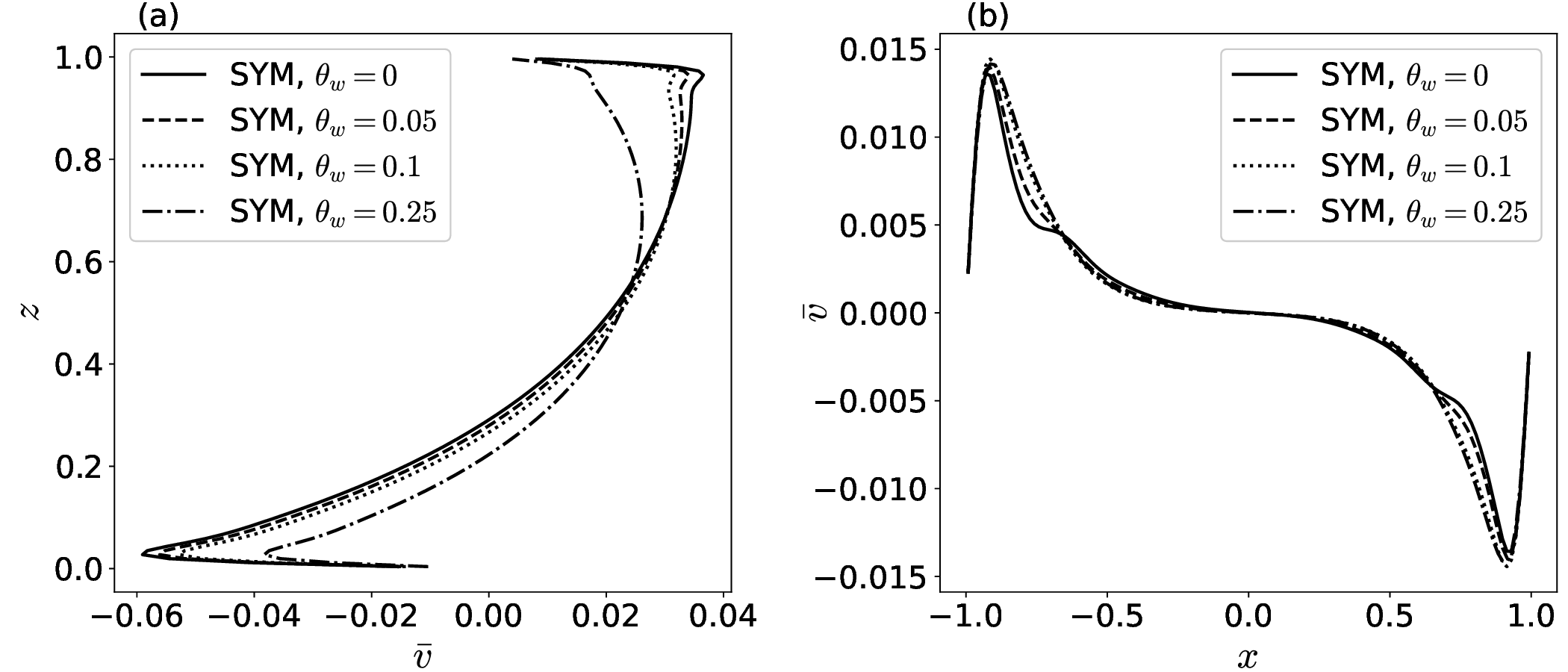}
\includegraphics[width=\columnwidth]{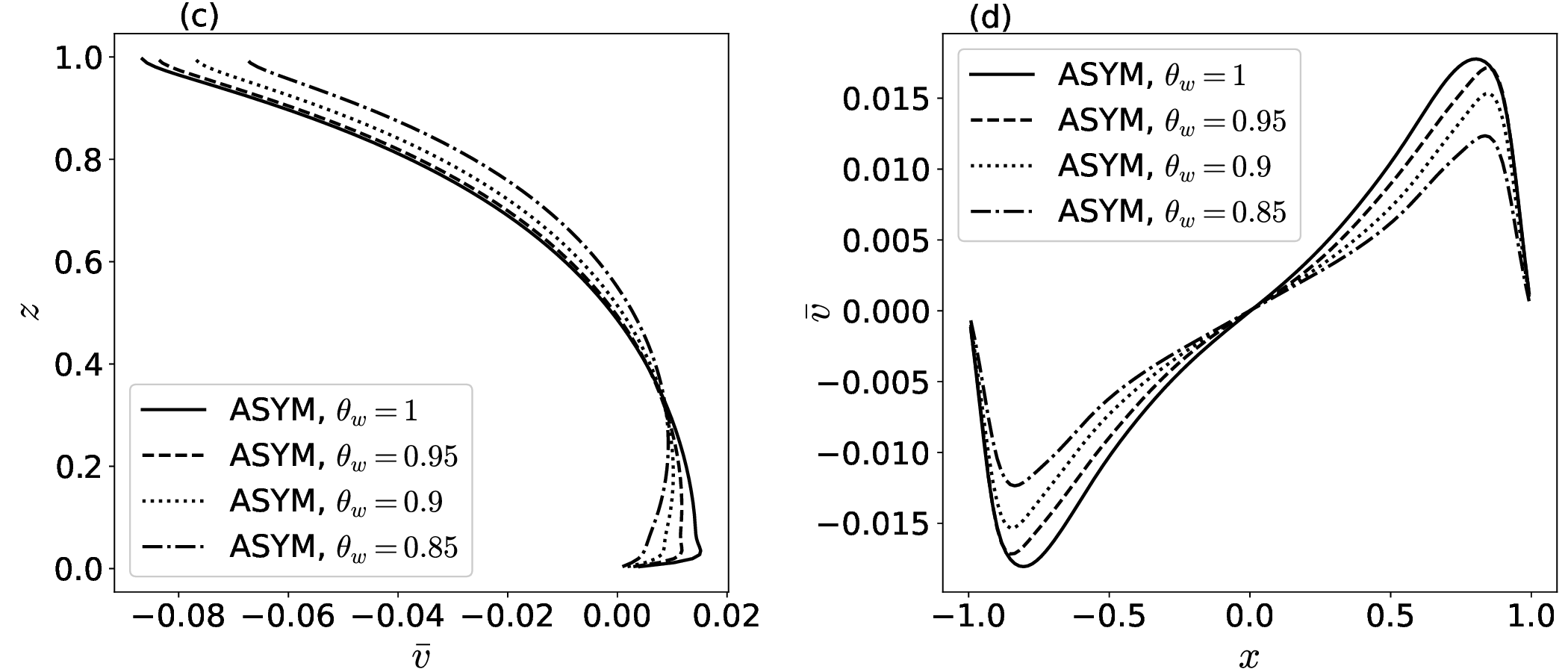}
\caption{\label{fig:tangvel_wallmodes_vs_rolls} {The tangential velocity as a function of (a,c) $z$ and (b,d) $x$ for cases where the flow takes the form of rolls and wall modes for (a,b) SYM and (c,d) ASYM BCs with fixed wall temperatures $\theta_w$. The combination of the depth-averaged tangential velocity and the vertical shear dictate the onset of wall modes from the steady roll state. Flow parameters are as in Figs. \ref{fig:tangvel_SYMNS} and \ref{fig:tangvel_ASYM}.}}
\end{figure}

{Figure \ref{fig:tangvel_wallmodes_vs_rolls}(a,b) shows that for SYM BCs the depth-averaged tangential velocity is similar for $\theta_w=0.25$ (steady rolls) and $\theta_w \leq 0.1$ (wall modes). The onset of wall modes for $\theta_w \leq 0.1$ is controlled by the vertical shear, which increases with decreasing $\theta_w$, rather than by the depth-averaged velocity. For ASYM BCs, the average tangential velocity is prograde, and the vertical shear increases as $\theta_w \rightarrow 1$.}
{This argument is bolstered by an examination of the case with the linear wall temperature profile (Eq. \ref{eq:lin_temp}), wherein the wall thermal forcing is minimal \citep[see][]{Ning1993}, and thus so too is the vertical shear. Therefore, the net tangential velocity is retrograde and the wall modes propagate in the retrograde direction for both SYM and ASYM velocity BCs.  These arguments apply in both cuboidal and cylindrical geometries.}

\begin{figure}
\centering
\includegraphics[width=\columnwidth]{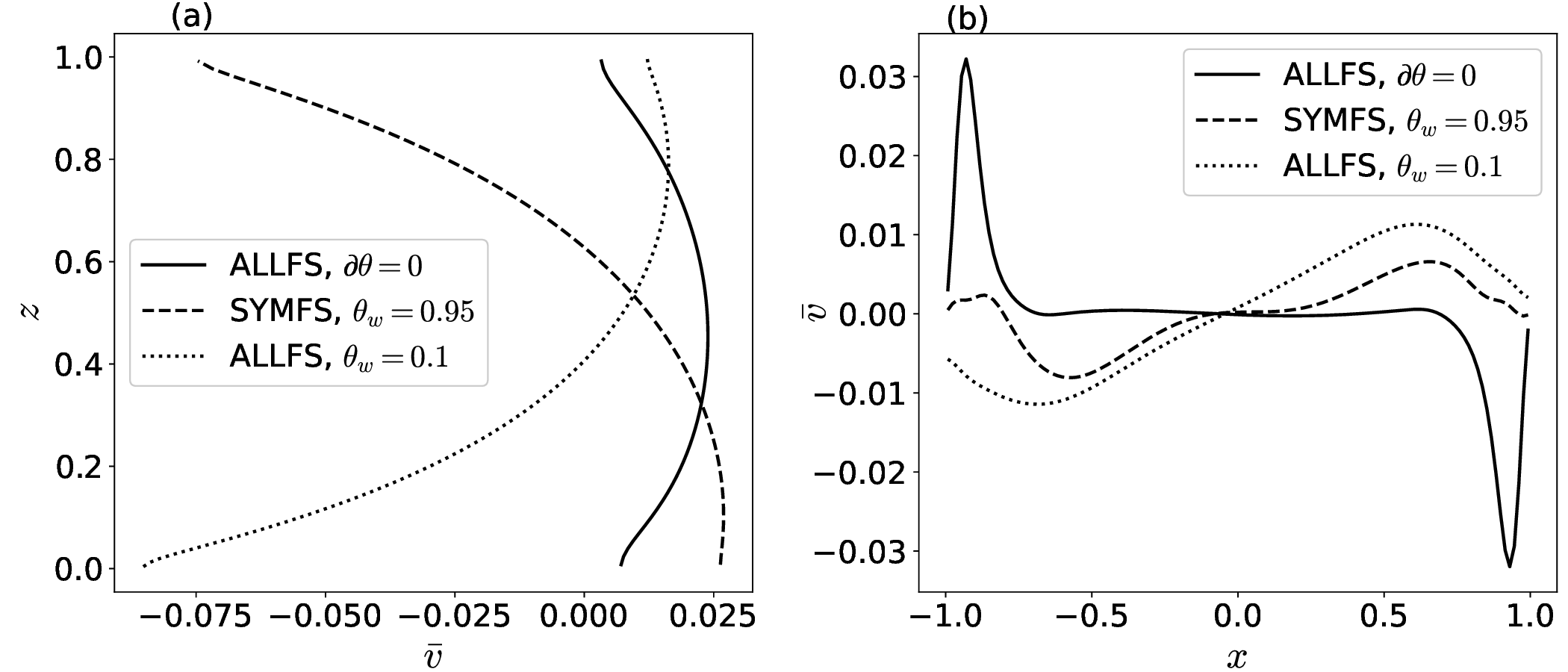}
\caption{\label{fig:tangvel_ALLFS} {As in Figures \ref{fig:tangvel_SYMNS} and \ref{fig:tangvel_ASYM}, but with SYMFS and ALLFS velocity BCs. Despite the large tangential velocities at the upper and lower boundaries seen in (a), the depth-averaged velocities in (b) are of much smaller magnitudes. }}
\end{figure}

{For ALLFS and SYMFS BCs, wall modes have no preferred direction of travel. In Fig. \ref{fig:tangvel_ALLFS} we see that, compared to the case with insulating walls, fixed-temperature walls lead to much smaller depth-averaged tangential velocities of indeterminate direction. The latter point is demonstrated by comparing the two different realizations of the ALLFS BCs.}

\begin{figure}
\centering
\includegraphics[width=\columnwidth]{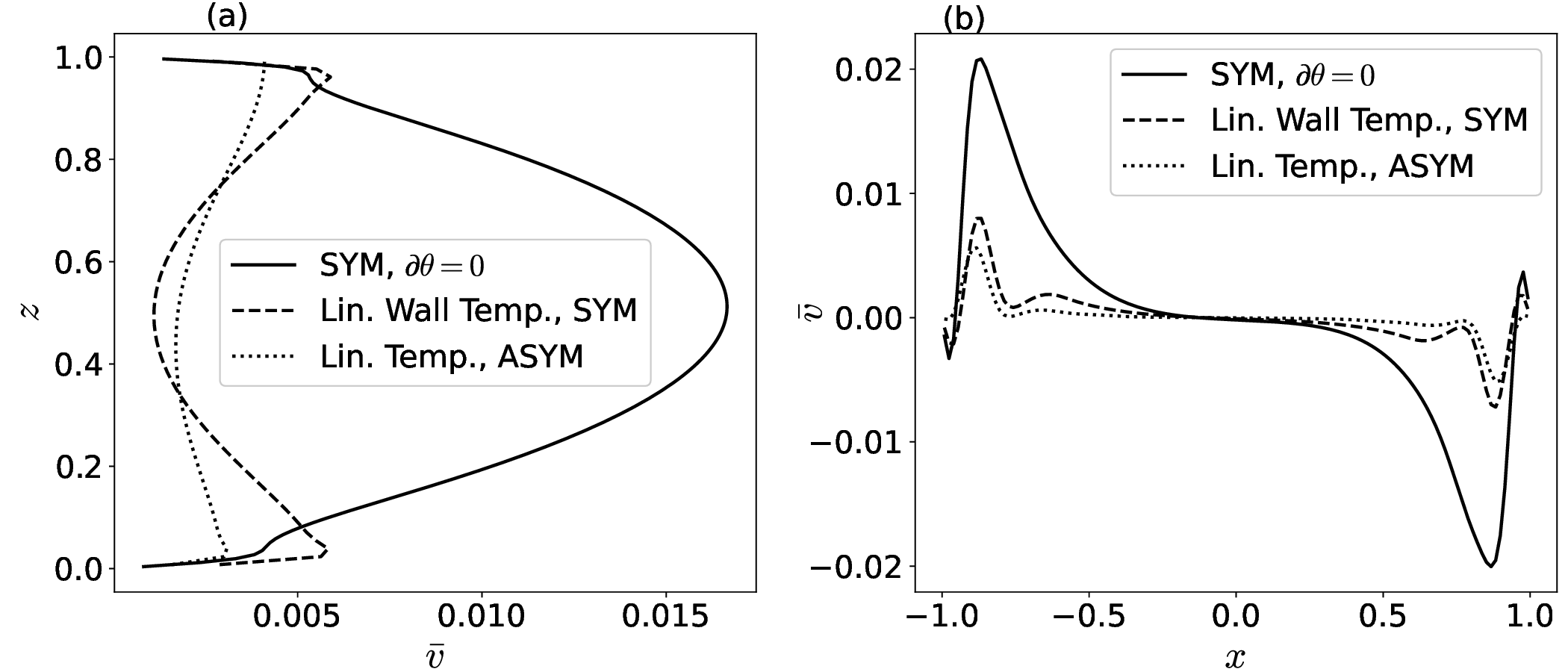}
\caption{{\label{fig:lintemp} Tangential velocity profiles versus (a) $z$, and (b) $x$, for the cases with the linear wall temperature of Eq. \ref{eq:lin_temp}.}}
\end{figure}

{Lastly, Fig. \ref{fig:lintemp} shows the tangential velocity profiles obtained with the linear wall temperature (Eq. \ref{eq:lin_temp}), no-slip BCs on the lower boundary at $z=0$ and either no-slip or {stress-free} BCs on the upper boundary at $z=1$. Since the thermal forcing at the wall is smaller than for fixed $\theta_w$, the vertical shear generated is negligible, and the average tangential velocity is retrograde. The wall modes propagate in a retrograde direction for both velocity BCs.}

\subsection{The influence of $\theta_w$ on wall mode formation} \label{sec:thetaw}

{From the results presented in \S \ref{sec:SYMNS_ASYM} and \S \ref{sec:other_BCs}, we 
see that wall modes only occur for values of $\theta_w$ 
sufficiently close to the upper or lower boundary temperatures, with the 
flow taking the form of steady rolls for $\theta_w > \thcone$ (see figure 
\ref{fig:wall_modes_symns}(a,c)). To determine the wall temperature $\thcone$ below which wall modes
exist, $\theta_w < \thcone$, we quantify the wall mode strength using the oscillation amplitude of the temperature $\theta (x,y,z,t)$ as}
\be
{A_\theta^2 = \langle \left(\theta - \bar{\theta} \right) ^ 2 \rangle,} \label{eq:amplitude}
\ee
{where $\bar{\theta}\left(x,y,z\right)$ is the time-averaged temperature at a
given location $(x,y,z)$, and the angle brackets $\langle \cdot 
\rangle$ denote an average over the spatial domain. Thus, the amplitude 
$A_\theta=0$ for steady (or zero) flow, and the wall temperature
$\theta_w$ at which $A_\theta > 0$ is the critical wall temperature 
$\thcone$. Similarly, the temperature $\thctwo$ is defined such that when $\theta_w > \thctwo$ wall modes occur, and when 
$\theta_w < \thctwo$, steady convective rolls occur. }

\begin{figure}
\centering
\includegraphics[width=0.8\columnwidth]{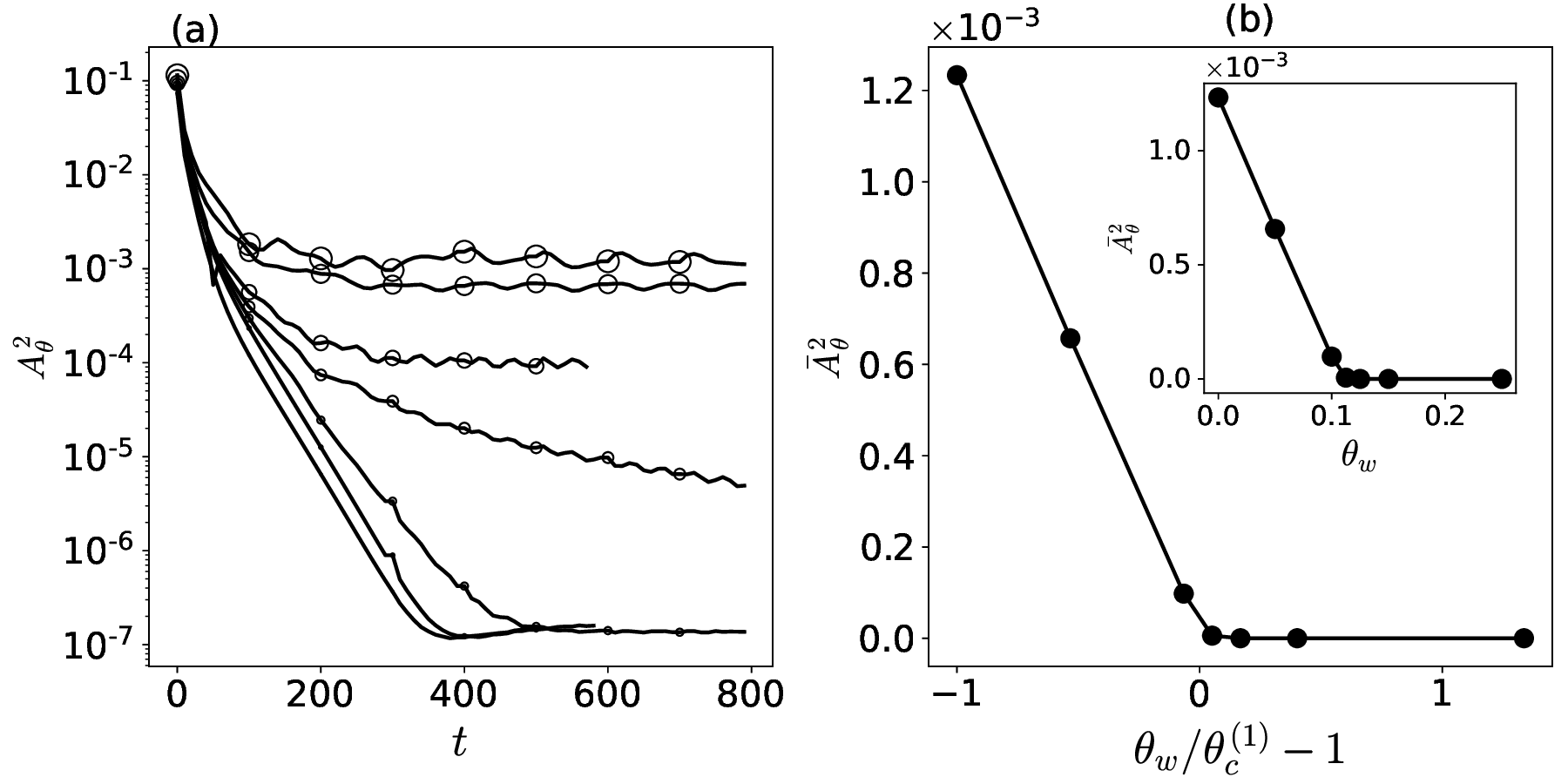}
\caption{\label{fig:amplitude_symNS} (a) The amplitude $A_\theta^2$ of the wall modes as a function of time, showing the existence of a steady oscillatory state, and (b) the domain-averaged oscillation amplitude $\bar{A}_\theta^2$ as a function of the reduced wall temperature $\theta_w / \thcone - 1$ for SYMNS BCs (see Table \ref{tab:BCs}) and fixed flow parameters $E=10^{-4}, Ra=10^6$, and $Pr=1$. In the inset of (b), $A_\theta^2$ is plotted versus the wall temperature $\theta_w$. The curves in (a) are plotted for the wall temperatures $\theta_w = (0, 0.05, 0.1, 0.1125, 0.125, 0.15, 0.25)$, {with circles of larger size for smaller $\theta_w$. The averages in (b) are obtained over the last $100$ flow units of each curve in (a).} The linear dependence of the (squared) amplitude on the magnitude of the deviation from the critical wall temperature is characteristic of a Hopf bifurcation.}
\end{figure}

{In figure \ref{fig:amplitude_symNS}, we plot the oscillation amplitude
$A_\theta$ as the wall temperature $\theta_w $ is varied for SYMNS BCs and  fixed flow parameters $E=10^{-4}, Ra=10^6$, and $Pr=1$. We find that the domain-averaged oscillation amplitude $\bar{A}_\theta^2 \propto |\theta_w-\thcone|$, with a critical wall temperature $\thcone \approx 0.1$. This dependence of the amplitude on the deviation from threshold is similar to the Hopf bifurcation that occurs for wall modes with insulating walls \citep{Ecke1992}, in which the controlling parameter is the Rayleigh number.}

\begin{figure}
\centering
\includegraphics[width=0.8\columnwidth]{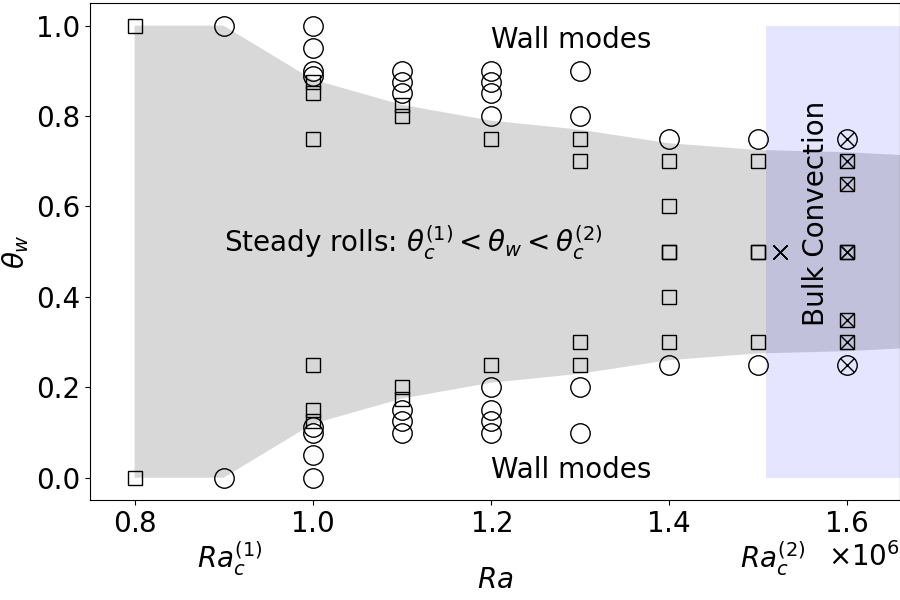}
\caption{\label{fig:cases} The different regimes of behavior  observed with conducting walls and SYMNS BCs. {The symbols denote whether wall modes (circles), steady rolls (squares) or bulk convection with columnar vortices (crosses) are seen. Note that combinations of these are possible. The regions shaded grey and blue correspond to the steady roll- and bulk-convective states. The lower and upper boundaries of the grey region correspond to the critical temperatures $\thcone$ and $\thctwo$ respectively.} For small Rayleigh numbers $Ra$, wall-attached steady rolls are seen for all $\theta_w$. For Rayleigh numbers greater than a critical value $Ra_c^{(1)}$, wall modes are seen for $\theta_w < \theta_c^{(1)}$ and $\theta_w > \theta_c^{(2)}$, while the steady roll state persists for $\theta_c^{(1)} < \theta_w < \theta_c^{(2)}$. For larger Rayleigh numbers, convection also sets in away from the walls, with the flow taking the form of a series of nested rolls that remain stable for extended periods of time. For $Ra > \Ractwo\equiv\Racb$ of Eq. \eqref{eq:Rac_bulk}, flow {in the bulk} takes the form of horizontally drifting columnar vortices that are typical of \RRBCalt, {while either nested rolls or wall modes are seen adjacent to the walls.} For $E=10^{-4}, Pr=1$, we find $\Racone=9\times10^5$, and $\Racb=1.5\times10^6$.}
\end{figure}

We repeat this exercise for different values of $Ra$, keeping
the parameters $E=10^{-4}$ and $Pr=1$ fixed. The smallest value of $Ra$ such
that $\thcone(Ra) > 0$ is defined as $\Racone=\Racone(E,Pr,\mathrm{BCs})$. {The 
resulting behavior for the cases we have simulated is summarized schematically in figure \ref{fig:cases}, where we find a monotonic dependence of $\thcone$ on $Ra$.}
{For SYMNS BCs and wall temperatures $\theta_w < \thcone$ and $\theta_w > \thctwo$ retrograde wall
modes are observed as seen in figure \ref{fig:Hovmoller_SYMNS}(a) and (b) respectively. 
The difference $\thctwo-\thcone$ decreases monotonically as $Ra$ increases.}\\

\begin{figure}
\centering
\includegraphics[width=\columnwidth]{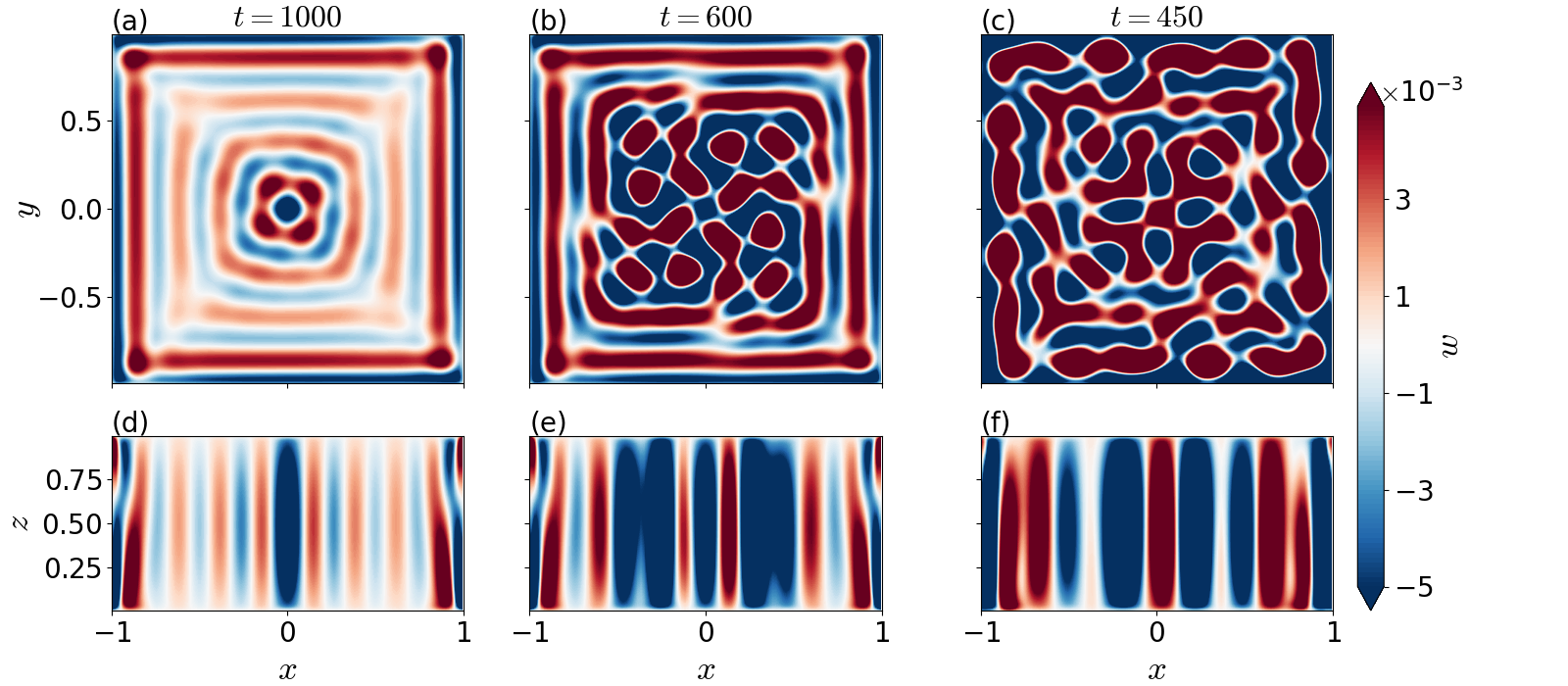}
\caption{\label{fig:steady_nested_rolls} {With $E=10^{-4}$, $Pr=1$, $\theta_w=0.4$ and SYMNS BCs, we see (a,d) for $Ra=\Racb=1.5\times10^6$, steady nested rolls spanning the entire horizontal extent of the domain; (b,e) wall-adjacent steady rolls with columnar vortices in the bulk for $Ra=1.6\times10^6$. Note that for $\theta_w=0.25 < \thcone$, we see (c,f) bulk convection coexist with retrograde propagating wall modes. See also Fig. \ref{fig:cases}.}}
\end{figure}

{As the Rayleigh number approaches the critical value for the onset of bulk flow, $\Racb=1.5\times10^6$ from Eq. \eqref{eq:Rac_bulk}, the difference $\thctwo-\thcone$ can either (a) vanish before bulk flow sets in, such that $\Ractwo < \Racb$, or (b) remain finite until bulk convection sets in at $\Ractwo = \Racb$. We find the latter case, wherein as $Ra$ increases the flow takes the form of a series of nested convection rolls that span the entire horizontal area of the domain, as shown in figures \ref{fig:steady_nested_rolls}(a,d).}
{For $Ra > \Racb$, as noted above, these rolls break down into individual columnar vortices in the fluid bulk, as seen in figure \ref{fig:steady_nested_rolls}(b,e), due to the mechanism reported by \citet{Boubnov1986,Zhong2010} and \citet{Ravichandran2020_RotCon}, into a state of geostrophic convection in the bulk. For larger $Ra$, the wall modes become less prominent.}

\begin{figure}
\centering
\includegraphics[width=1.0\columnwidth]{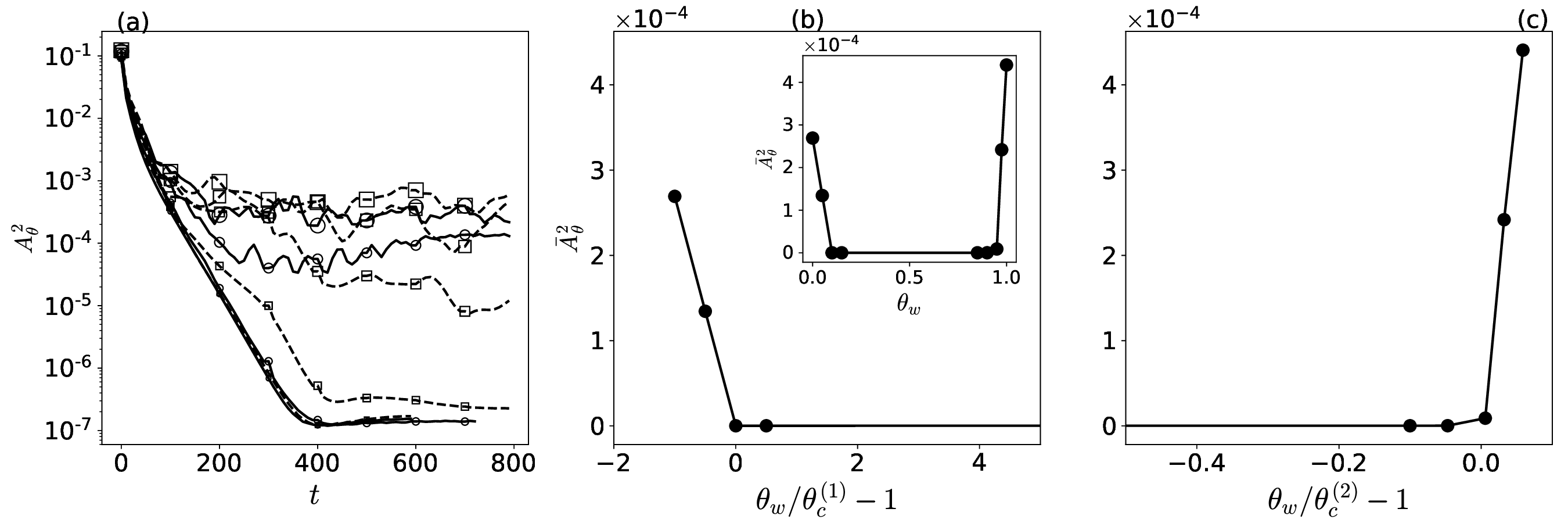}
\caption{\label{fig:amplitude_asymFS} (a) The amplitude $A_\theta^2$ of the spatio-temporal oscillations as a function of time, showing the existence of a steady oscillatory state, and the domain-averaged oscillation amplitude $\bar{A}_\theta^2$ as a function of (b) the reduced wall temperature $\theta_w / \thcone - 1$, and (c) $\theta_w/\thctwo - 1$, for ASYM BCs (see Table \ref{tab:BCs}). In the inset of (b), $A_\theta^2$ is plotted versus the wall temperature $\theta_w$.  The curves in (a) are plotted for the wall temperatures $\theta_w = (0, 0.05, 0.1, 0.15)$ (solid lines, circles of decreasing size) and $\theta_w= (0.85, 0.9, 0.95, 0.975, 1.0)$ (dashed lines, squares of increasing size). The averages in (b) are obtained over the last 100 flow units for each curve in (a). The linear dependence of the (squared) amplitude on the magnitude of the deviation from the critical wall temperature is characteristic of a Hopf bifurcation. Note the asymmetry between $\thcone$ and $\thctwo$ (c.f. figure \ref{fig:amplitude_symNS}). }
\end{figure}

The symmetry between $\thcone$ and $\thctwo$ observed for SYMNS BCs, where $\thcone + \thctwo = 1$, breaks
down if the velocity boundary conditions are asymmetric, such as for ASYM BCs.
In figure \ref{fig:amplitude_asymFS}, we plot the oscillation amplitude 
$A_\theta^2 (\theta_w)$ for the same parameters as in figure 
\ref{fig:amplitude_symNS}, but with ASYM BCs, showing that $\thcone + \thctwo > 1$.

{Of further relevance to figure \ref{fig:cases} is the fact that in the limit $E\rightarrow0$, equations \eqref{eq:RacwN} and \eqref{eq:RacwD} for the critical Rayleigh numbers are independent of the Prandtl number. For finite $E$, \cite{Herrmann1993} found that as the Prandtl number decreased so too did $Ra_c$ (see their figures 3 and 6), driving the system towards instability. Similarly, decreasing the Ekman number increases the critical Rayleigh number, and leads to the same qualitative effects as does decreasing $Ra$. Next we consider further the effects of varying the Prandtl number. }

\subsection{{The role of the Prandtl number \label{sec:Prandtl}}}

{An important effect of varying the Prandtl number to control the thickness of the thermal boundary layers at the walls and the resultant heat transfer. Thus, for a Boussinesq fluid, decreasing the Prandtl number is associated with increasing the thermal diffusivity.}

\begin{figure}
\centering
\includegraphics[width=0.6\columnwidth]{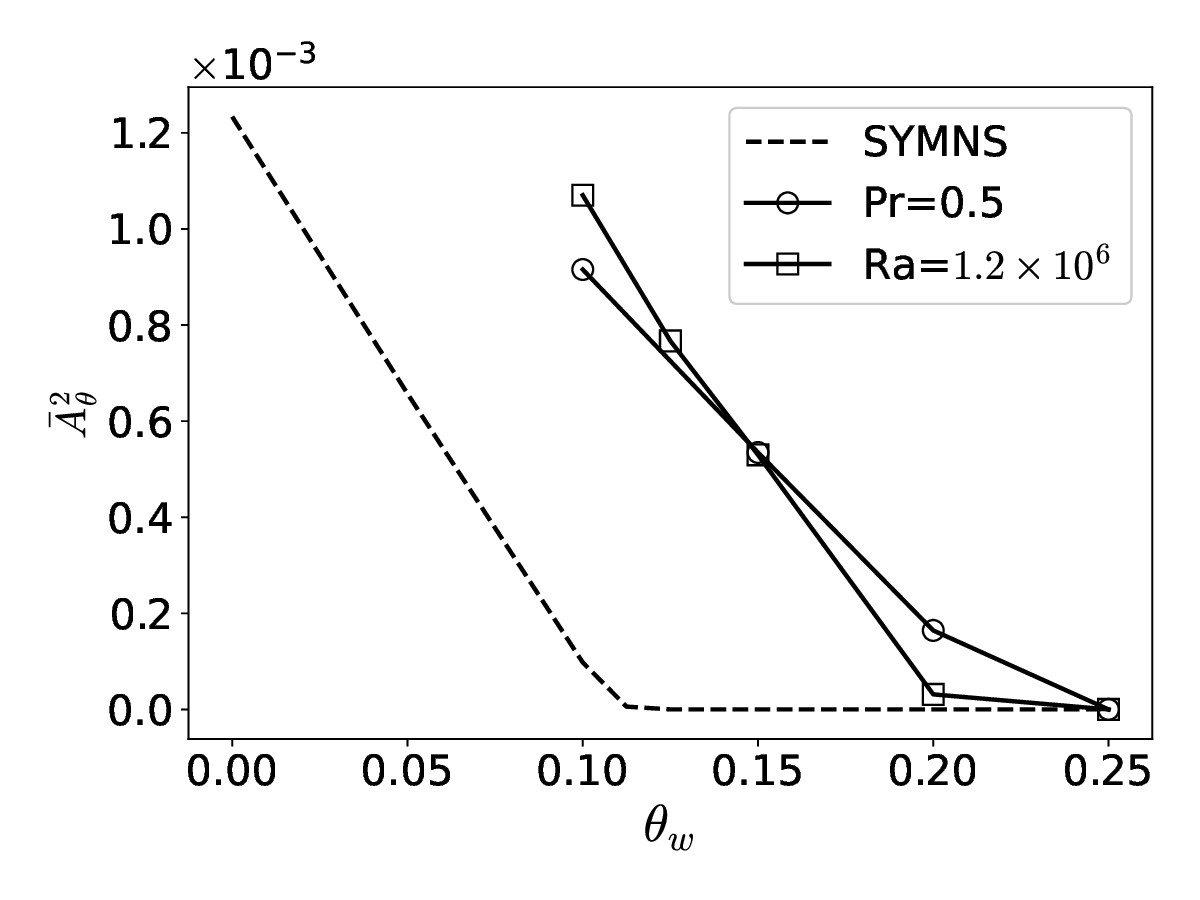}
\caption{\label{fig:ampl_vs_Ra_Pr}{Oscillation amplitudes with $E=10^{-4}$, as in Fig. \ref{fig:amplitude_symNS}(b), but with $Ra=10^6$ and $Pr=0.5$ (open circles), and $Ra=1.2\times10^6$, $Pr=1$ (open squares). The dashed curve is the case $E=10^{-4}$, $Ra=10^6$ and $Pr=1$ exactly as in Fig. \ref{fig:amplitude_symNS}(b).}}
\end{figure}

\begin{figure}
\centering
\includegraphics[width=0.8\columnwidth,trim={50 0 10 0},clip]{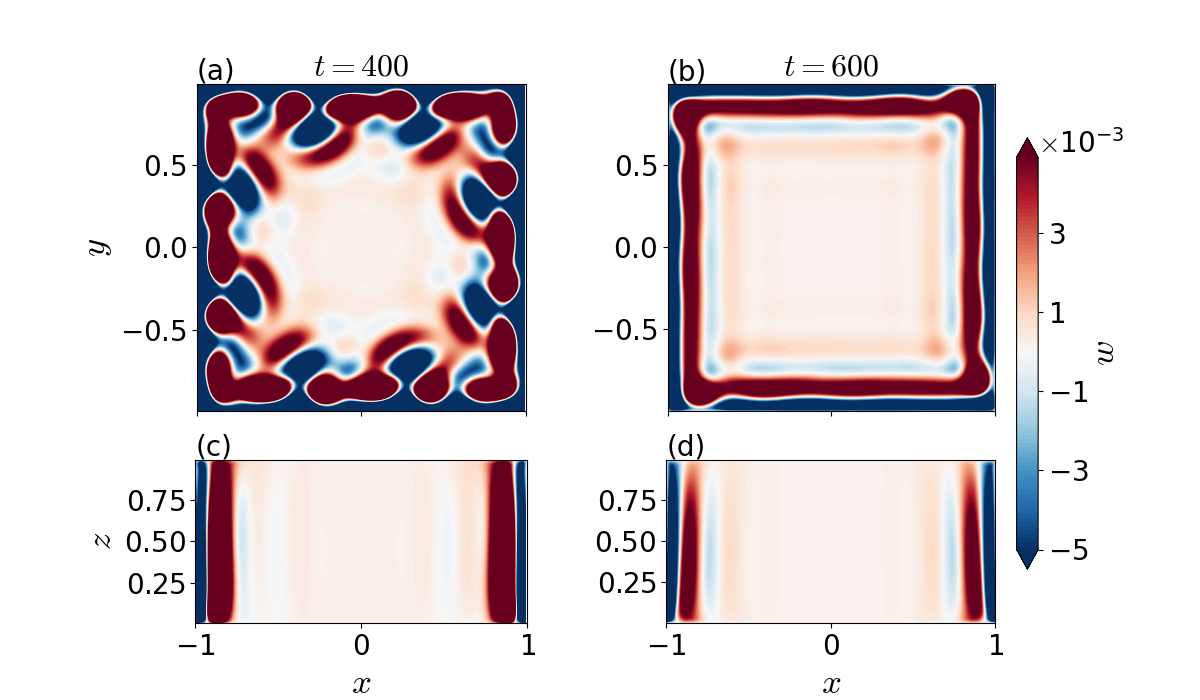}
\caption{{\label{fig:wall_modes_vs_Pr} With $E=10^{-4}$, $Ra=10^6$ and SYMNS BCs, we see wall modes for (a,c) $Pr=0.5$ and $\theta_w=0.15$; and a steady convective state with wall-adjacent rolls for (b,d) $Pr=2$ and $\theta_w=0$. The subplots show the horizontal (a,b) and vertical (c,d) cross-sections of the vertical velocity $w$.}}
\end{figure}

{Considering again the case of $\theta_w \approx 1$, with other parameters held constant, smaller Prandtl numbers result in larger buoyancy forcing at the walls, leading to larger wall-adjacent vertical velocities, and larger retrograde velocities as the flow turns inwards at the upper boundary. As a result, the flow experiences greater vertical shear, which results in the onset of wall modes for a smaller $\theta_w$.}
{We see in Fig. \ref{fig:ampl_vs_Ra_Pr} that decreasing $Pr$ and increasing $Ra$ both increase $\thcone$. Thus,}
{for $Pr = 0.5$, wall modes are seen for $\theta_w=0.15$ in Fig. \ref{fig:wall_modes_vs_Pr}(a), whereas steady rolls are seen for $Pr=1$.}
{In contrast, for $Pr=2$, the steady roll state is seen for $\theta_w=0$ in Fig. \ref{fig:wall_modes_vs_Pr}(b), and larger values of $Ra$ are needed for the onset of wall modes than for the onset of bulk convection. Therefore, smaller Ekman numbers, and thus more strongly rotation-dominated flows, are needed for wall modes when $Pr>1$.}

\section{Conclusion} \label{sec:conclusion}

The flow structure in confined \RRBC is comprised of alternating regions of upwelling warm and downwelling cold fluid. When conditions lead to these patterns being adjacent to the walls of the system, they are commonly referred to as `wall-modes', and were first observed in laboratory experiments \cite[See][and refs in the latter]{Rossby1969,Ecke1992}. Here, in geometries of aspect ratio greater than unity, we have used direct numerical simulations to study the formation and spatio-temporal evolution of wall-adjacent flow patterns when the walls are conducting instead of insulating. We showed that the wall temperature $\theta_w$ controls whether the flow takes the form of steady convective rolls or propagating wall-modes. We found that the velocity BCs are crucial to the dynamics of wall modes, and that these modes propagate in a fixed direction only if at least one of the upper and lower boundaries obeys the no-slip condition. Moreover, the direction of propagation of the wall modes can be reversed for suitable combinations of the velocity BCs at the upper and lower boundaries and the wall temperature.  In particular, if only the upper boundary is stress-free, the wall modes propagate in a prograde direction when $\theta_w \approx 1$. 
Indeed, there is a similarity between the effects of velocity BCs on wall mode formation and propagation, to transient convective ring formation \citep{Boubnov1986,Vorobieff1998,Zhong2010,Ravichandran2020_RotCon}.

{Although wall modes were previously shown to occur when the walls are conducting, we found that asymmetric velocity BCs drive an asymmetry in the torques exerted at the two boundaries.  Such effects may have astro- or geophysical consequences, where asymmetric velocity BCs are common, such as for example in natural bodies of water with free upper surfaces. The effects of the asymmetry in velocity BCs on integral flow properties, such as the helicity \citep{Moffatt1992} are a subject of ongoing study. }

{Finally, the rich range of flow behavior  in slightly supercritical Rayleigh-B\'enard convection has served as a model for the study of nonlinear-dynamical systems, such as the Benjamin-Feir and Eckhaus instabilities \citep{Ning1993,liuEckhausBenjaminFeirInstabilityRotating1997a,liuNonlinearTravelingWaves1999a,lopezCrossflowInstabilityFinite2009}.}
More recent studies have shown that the wall-modes in \RRBC {may be } a topologically conserved feature, robust to severe {vertically homogeneous} modifications to the geometry \citep[][]{Favier2020}. Therefore, the influence of conducting walls in combination with asymmetric velocity BCs on the wall modes may provide a framework of general interest in the theory of pattern formation.

\section*{Acknowledgements}
Computational resources from the Swedish National Infrastructure for Computing (SNIC) under grants SNIC/2020-5-471, SNIC/2021-5-449 and SNIC/2022-5-473 are gratefully acknowledged. Computations were performed on Tetralith. The Swedish Research Council, under grant no. 638-2013-9243, is gratefully acknowledged for support. Nordita is partially supported by Nordforsk. Since July 2022, SR is supported through Seed Grant RD/0522-IRCCSH0-020 from IIT Bombay. \tb{We thank the anonymous referees for helpful comments.}

\section*{Declaration of interests.} The authors report no conflict of interest.

%
    
\end{document}